\documentclass[aps,prl,reprint,amsmath,amssymb]{revtex4-2}

\usepackage{graphicx}\usepackage{dcolumn}\usepackage{bm}\usepackage{hyperref}

\usepackage{physics} \usepackage{tensor} \usepackage[dvipsnames]{xcolor}

\usepackage{pdfpages}
\makeatletter
\AtBeginDocument{\let\LS@rot\@undefined}
\makeatother

\newcommand{\bPsi}{{\bm{\mathsf{\Psi}}}}

\newcommand{\bx}{{\bm x}}

\newcommand{\bk}{{\bm k}}

\newcommand{\bw}{{\bm w}}

\newcommand{\bq}{{\bm q}}
\newcommand{\bv}{{\bm v}}

\newcommand{\bA}{{\bm A}}
\newcommand{\Schr}{Schr\"{o}dinger}

\begin{document}

\preprint{APS/123-QED}

\title{Generating Moving Field Initial Conditions with Spatially Varying Boost}

\author{Siyang Ling}
\affiliation{City University of Hong Kong,\\
  Tat Chee Avenue, Kowloon, Hong Kong SAR, China}
\email{siyaling@cityu.edu.hk}

\date{\today}

\begin{abstract}
  We introduce a novel class of algorithms, the ``spatially varying boost'', for generating dynamical field initial conditions with prescribed bulk velocities.
  Given (non-moving) initial field data, the algorithm generates new initial data with the given velocity profile by performing local Lorentz boosts.
  This algorithm is generic, with no restriction on the type of the field, the equation of motion, and can endow fields with ultra-relativistic velocities.
  This algorithm enables new simulations in different branches of physics, including cosmology and condensed matter physics.
  For demonstration, we used this algorithm to (1) boost two  Sine-Gordon solitons to ultra-relativistic speeds for subsequent collision, (2) generate a relativistic transverse Proca field with random velocities, and (3) set up a spin-$1$ \Schr-Poisson field with velocity and density perturbations consistent with dark matter in matter dominated universe.
\end{abstract}

\maketitle

\section{Introduction}
\label{sec:intro}

Accurate initial conditions are crucial for numerical simulations of any physical system.
Oftentimes, matter fields need to be initialized with prescribed velocities.
For example, in cosmological structure formation simulations, matter fields are initialized with velocities given by predictions from cosmological perturbation theory~\cite{Efstathiou:1985re,Matarrese:1992rp,Scoccimarro:1997gr,Valageas:2001qe,Yoshida:2003sy,Sirko:2005uz,Crocce:2006ve,Tatekawa:2007ix,Hahn:2011uy,LHuillier:2014fcj,Hahn:2014lca,Fidler:2015npa,Adamek:2015eda,Adamek:2016zes,Zennaro:2016nqo,Garrison:2016vvp,Tatekawa:2019zpc,Springel:2020plp,Hahn:2020lvr,Michaux:2020yis,Barrera-Hinojosa:2020arz,Angulo:2021kes,List:2023kbb}.
In reionization simulations, accurate modeling of initial velocities is crucial for obtaining accurate kinetic Sunyaev-Zel'dovich observables~\cite{Iliev:2005sz,Mesinger:2010ne,Vogelsberger:2014dza,Molaro:2019mew,Maity:2021mhr,Trac:2021qbn,Chen:2022lhr,Gnedin:2022eza}.
Simulating collisions of solitons, kinks and interconnections of strings also requires boosted initial conditions~\cite{Sugiyama:1979mi,Matzner:1988qqj,Lee:1991ax,Manton:2004tk,Vachaspati:2006zz,Achucarro:2006es,Dorey:2011yw,Gani:2014gxa,Amin:2014fua,Amin:2013dqa,Amin:2013eqa,Kinach:2024hfa,Kinach:2024qzc}.
In quark-gluon plasma simulations, heavy ions are boosted to relativistic velocities to resemble their high-energy collisions in colliders~\cite{Vredevoogd:2009zu,Schenke:2012wb,Gale:2013da,Heinz:2013th,Moreland:2014oya,Carrington:2020ssh}.
It is thus useful to have generic numerical methods for assigning velocities to matter fields.

Generating moving initial conditions is conceptually straightforward in most types of simulations.
For particle (e.g., N-body, SPH) and fluid based methods (e.g., finite volume methods), one simply assigns velocities to individual particles or fluid elements~\cite{Joyce:2004em}; for phase-space methods (e.g., Vlasov equations), one generates anisotropic velocity distributions $f(\bv)$~\cite{Amin:2025dtd}.
In contrast, for dynamical field equations (e.g., wave equation), velocity is not fundamental but derived, either from momentum and energy densities ($\bv = \bq / \rho$) or via the Madelung transform ($\bv = \nabla S / m$)~\cite{Madelung:1927ksh}.
These definitions have convoluted dependence on the field and its spacetime derivatives, and thus provide little insight on how to generate fields with prescribed bulk velocities.
Consequently, generating moving initial conditions for dynamical fields presents unique challenges in the landscape of numerical simulations.

Existing methods for generating moving field initial conditions typically require analytic solutions to the field equations~\cite{Matzner:1988qqj,Shellard:1987bv,Achucarro:2006es}, or utilize ansatz constructions~\cite{Veltmaat:2018dfz,Figueroa:2020rrl,May:2022gus,Kinach:2024hfa,Kinach:2024qzc}.
These methods can only generate a restrictive class of initial conditions.
In general, a field with arbitrary microscopic behavior can possess independent bulk motion, and a robust methodology should enable independent control over both aspects.
For instance, a photon fluid with bulk motion exhibits microscopic oscillations in velocity $\bq / \rho$ at the scale of the photon wavelength, while the bulk velocity operates on a much larger scale.

In this Letter, we introduce a novel class of algorithms called ``spatially varying boost'', which is given by eq.~\eqref{eq:infinitesimal_boost} and associated discussions.
We propose to generate moving field configurations by performing ``spatially varying boost'' on suitable non-moving fields, such that the fields appear to be boosted to velocity $\bv(\bx)$ around location $\bx$.
In particular, ``spatially varying boost'' reduces to an ordinary Lorentz boost in the limit of spatially uniform velocity $\bv(\bx)$.

Our algorithm resolves the aforementioned difficulties on assigning field velocities and enables novel simulations across cosmology, condensed matter, and high-energy physics.
By choosing appropriate field and velocity profiles, one can generate initial data for soliton collision/deformation, cosmic string bending, hot cosmological fields with bulk motion, etc.
Many of these initial conditions cannot be attained by existing methods.
To demonstrate the algorithm's validity and potential use, we discuss three implemented numerical examples in this Letter.

We refer to a general Lorentz covariant field theory specified by Lagrangian density $\mathcal{L}(x^\mu, F_a, \partial_\mu F_a)$, consisting of dynamical fields $F_a(x^\mu)$ indexed by $a$.
We use the mostly plus metric sign convention.
The stress-energy tensor contributed by the fields is $T\indices{^\mu_\nu} \equiv -2 (\sqrt{-g})^{-1} \fdv*{ (\int \sqrt{-g} \mathcal{L} \dd[4]{x} ) }{g\indices{^\mu^\nu}}$.
The energy and momentum densities are defined as $\rho \equiv -T\indices{^0_0}$ and $\bq \equiv T\indices{^0_i} {\bm e}^i$.
Einstein summation convention applies for field index $a$.

\section{Spatially Varying Boost}
\label{sec:procedure}

We first review the procedure for boosting a field by spatially uniform velocity $\bv$: simply view the field in a boosted frame of velocity $-\bv$.
Let $\Lambda\indices{^\mu_\nu}$ be the Lorentz transform with velocity $-\bv$, and $x^{\prime\mu} = \Lambda\indices{^\mu_\nu}x^\nu$ be the boosted coordinates.
The boosted fields $\hat{F}_a$ are given by:
\begin{align}
  \hat{F}_a(x') &= M_{ab}(\Lambda) F_b(\Lambda^{-1} x') \nonumber \\
  \partial_\mu' \hat{F}_a(x')
                &= M_{ab}(\Lambda) (\Lambda^{-1})\indices{^\nu_\mu}  \partial_\nu F_b( \Lambda^{-1} x' )  \,,
                \end{align}
where $\partial_\mu'$ are derivatives with respect to $x^{\prime\mu}$, and $M_{ab}$ is a representation of the Lorentz group.
The boosted field initial conditions are then $F_a(x')$ and $\partial_t' F_a(x')$ evaluated at $t'=0$:
\begin{align}
  \label{eq:uniform_velocity_boosted_ic}
  \hat{F}_a(0,\bx') &= M_{ab} F_b\left(- \gamma \bv \cdot \bx', \bx' + (\gamma - 1)\frac{\bv \cdot \bx'}{v^2}\bv \right) \nonumber \\
  \partial_t' \hat{F}_a(0,\bx') &= M_{ab} \gamma \left( \partial_t F_b(\Lambda^{-1}x') - \bv \cdot \nabla F_b(\Lambda^{-1}x') \right) \,,
\end{align}
where $ \gamma = 1 / \sqrt{1-v^2}$ and $\nabla = {\bm e}^i \partial_i$.

\begin{figure}[t]
  \centering
  \includegraphics[width=0.5\textwidth]{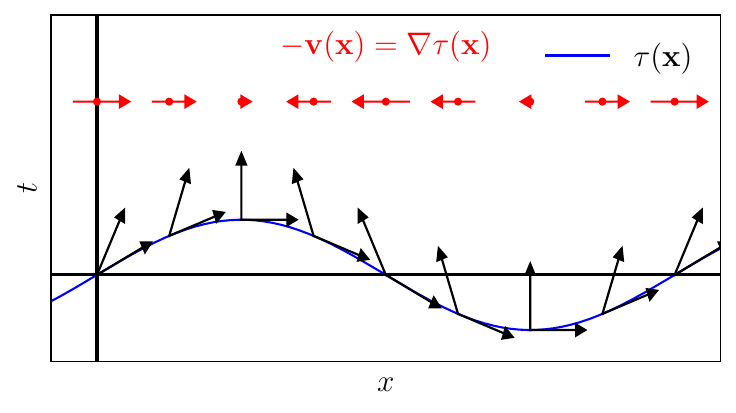}
  \caption{Spatially varying boost for $v \ll 1$. The black arrows are the axes for the locally boosted Lorentz frames. The arrows for spatial axes are tangent to the curve $t = \tau(x)$. The red arrows are the velocities $-\bv(\bx)$ of the respective Lorentz transforms.}
  \label{fig:illustration_2}
\end{figure}

Using the ordinary Lorentz boost for intuition, we introduce a spatially varying boost of bulk velocity $-\bv(\bx)$ for $v \ll 1$.
Note that for ordinary boosts, the $t' = 0$ time slice is given by the $t = - \bv \cdot \bx$ hyperplane.
In analogy, we define a ``constant time slice'' function $\tau(\bx)$ satisfying $\nabla \tau(\bx) = - \bv(\bx)$.
The graph of function $\tau(\bx)$ maps out a spacetime hypersurface, and its tangent hyperplanes are identified with the constant time slice of a Lorentz boosted frame centered at $(\tau(\bx), \bx)$.
See fig.~\ref{fig:illustration_2} for illustration.
Neglecting $\order{v^2}$ differences, we are motivated by eq.~\eqref{eq:uniform_velocity_boosted_ic} to define the following locally boosted field initial conditions:
\begin{align}
  \label{eq:varying_velocity_boosted_ic}
  \hat{F}_a(0,\bx) & = M_{ab}(\bx) F_b(\tau(\bx), \bx ) \nonumber \\
  \partial_t \hat{F}_a(0,\bx) &= M_{ab}(\bx) \gamma(\bx) ( \partial_t F_b(\tau(\bx), \bx ) \nonumber \\
  &\qquad - \bv \cdot \nabla F_b(\tau(\bx), \bx ) )
\end{align}
where $M_{ab}(\bx)$ and $\gamma(\bx)$ are the representation and gamma factor for the boost of velocity $- \bv(\bx)$.

Does the initial condition defined in Eq.~\eqref{eq:varying_velocity_boosted_ic} yield the same stress-energy tensor as $T\indices{^\mu_\nu}(\tau(\bx),\bx)$ boosted by $-\bv$?
To answer this question, note that $\hat{T}\indices{^\mu_\nu}$ is a Lorentz equivariant function of $\hat{F}_a$ and $\partial_\mu \hat{F}_a$; under a Lorentz transform of $\hat{F}_a(0,\bx)$ and $\partial_\mu \hat{F}_a(0,\bx)$, $\hat{T}\indices{^\mu_\nu}(0,\bx)$ undergoes the same Lorentz transform.
By construction, $\hat{F}_a(0,\bx)$ and $\partial_t \hat{F}_a(0,\bx)$ are identical to $M_{ab} F_b$ and $M_{ab} (\Lambda^{-1})\indices{^\nu_t} \partial_\nu F_b$, consistent with a Lorentz transform $\Lambda(-\bv(\bx))$.
On the other hand, direct computation shows $\partial_i \hat{F}_a(0,\bx) = M_{ab} (\Lambda^{-1})\indices{^\nu_i} \partial_\nu F_b + (\partial_i M_{ab}) F_b + \order{v^2}$.
Therefore:
\begin{align}
  \label{eq:varying_velocity_boosted_stress_energy}
  \hat{T}\indices{^\mu_\nu}(0,\bx) &=  \Lambda\indices{^\mu_\alpha} (\Lambda^{-1})\indices{^\beta_\nu} T\indices{^\alpha_\beta}(\tau(\bx),\bx) \nonumber \\ 
                                   &\qquad + \order{v^2} + \order{(\nabla M_{ab})F_b} \,.
\end{align}
Here, $(\nabla M_{ab})F_b = \order{\nabla \bv} M_{ab} F_b$ is negligible when $\nabla \bv / v \ll \nabla F_a / F_a$, or in the limit of spatially uniform $\bv(\bx)$.

Eq.~\eqref{eq:varying_velocity_boosted_ic} already gives a recipe for spatially varying boosts of velocity $\bv(\bx) \ll 1$.
We can further extend it to relativistic $\bv(\bx) \sim 1$ by composing infinitesimal boosts of $\bv(\bx) \ll 1$.
To this end, we introduce a book keeping parameter $\epsilon$, and redefine $ \tau(\bx) \to \epsilon \tau(\bx) $ and $\bv(\bx) \to \epsilon \bv(\bx)$.
The infinitesimal spatially varying boost can be identified by linearizing eq.~\eqref{eq:varying_velocity_boosted_ic} with respect to $\epsilon$ at $\epsilon = 0$.
Given $\tau(\bx)$, we define a collection of initial data $(\tilde{F}_a(\bx), \partial_t \tilde{F}_a(\bx))$ labeled by $\epsilon$, by the initial value problem:
\begin{align}
  \label{eq:infinitesimal_boost}
  & \eval{\tilde{F}_a(\bx)}_{\epsilon = 0} = F_a(0, \bx),\quad
                       \eval{\partial_t\tilde{F}_a(\bx)}_{\epsilon = 0} = \partial_t F_a(0, \bx)   ,                  \nonumber \\
  & \dv{\epsilon} \tilde{F}_a(\bx) = \tau(\bx) \partial_t\tilde{F}_a(\bx) +  N_{ab}(\bx) \tilde{F}_b(\bx) \nonumber \\
  & \dv{\epsilon} \partial_t \tilde{F}_a(\bx) = \tau(\bx)  \partial_t^2 \tilde{F}_a(\bx )  + \nabla \tau(\bx) \cdot \nabla \tilde{F}_a(\bx ) \nonumber \\
  &\qquad \qquad \qquad + N_{ab}(\bx)  \partial_t \tilde{F}_b(\bx ) \, , \nonumber \\
  &\textrm{where } N_{ab}(\bx) \equiv \eval{ \dv{\epsilon}  M_{ab}(\Lambda(\epsilon\nabla\tau(\bx))) }_{\epsilon = 0} \,,
\end{align}
and the equation of motion gives $\partial_t^2 \tilde{F}_a(\bx)$ as a functional of $\tilde{F}_a$ and $\partial_t\tilde{F}_a$.
The boosted initial conditions are given by $(\tilde{F}_a(\bx), \partial_t \tilde{F}_a(\bx))$ at $\epsilon = 1$.
This initial value problem can be readily solved numerically via time-stepping algorithms.

In the boost defined by eq.~\eqref{eq:infinitesimal_boost}, $-\nabla \tau(\bx)$ should be interpreted as the spatially varying rapidity $\bw(\bx) = -\nabla \tau(\bx) = \tanh^{-1}(v) \bv / v$, instead of velocity $\bv(\bx)$.
This is because rapidity is the affine parameter associated with infinitesimal Lorentz boosts.
To see this, note that taking $\tau(\bx) = - \bw \cdot \bx$ for constant $\bw$ gives exactly the ordinary Lorentz boost with rapidity $\bw$.
The algorithm is thus valid for $w > 1$, and reduces to eq.~\eqref{eq:varying_velocity_boosted_ic} for $w \approx v \ll 1$.

Due to Lorentz length contraction, for relativistic boosts $v \sim 1$, the boosted stress-energy $\tilde{T}\indices{^\mu_\nu}(\bx)$ is no longer a Lorentz transform of $T\indices{^\mu_\nu}(0,\bx)$ as in eq.~\eqref{eq:varying_velocity_boosted_stress_energy}.
For a boost with $\gamma \gg 1$, the energy density $\rho \equiv - T\indices{^0_0}$ transforms as $\rho \to \gamma^2 \rho$, and length contracts by $1 / \gamma$.
These two effects compensate to yield a transform $E \to \gamma E$ for the total energy, as expected for a particle.
Simply performing a local Lorentz transform on $T\indices{^\mu_\nu}(0,\bx)$ does not account for length contraction, and hence does not give the correct $\tilde{T}\indices{^\mu_\nu}(\bx)$.
Nevertheless, due to cancellation of $\gamma$ factors in $\bq$ and $\rho$, the bulk velocity $\bq(\bx) / \rho(\bx)$ of the boosted field is still expected to be $\sim \bv(\bx)$.
Numerical experiments also support this claim.

\section{Random fields}
\label{sec:random_fields}
Of particular interest is the case where the field being boosted is a random field described by an ensemble.
For statistically homogeneous and isotropic field ensembles, the expected stress-energy tensor is of the form $\expval{T\indices{^\mu_\nu}(t,\bx)} = \mathrm{diag}(-\bar{\rho}, \bar{p}, \bar{p}, \bar{p})$, resembling a non-moving fluid.
For non-relativistic boosts ($v \ll 1$), eq.~\eqref{eq:varying_velocity_boosted_stress_energy} gives the stress-energy of the boosted fields:
\begin{align}
  \expval{\tilde{T}\indices{^\mu_\nu}(\bx)}
 &= \left[ (\bar{\rho} + \bar{p}) u^\mu u_\nu  + \bar{p} \delta\indices{^\mu_\nu} \right] \left(1 + \order{\frac{\nabla \bv / v}{\nabla F_a / F_a}}\right)
   \end{align}
where $u^\mu = \gamma (1,\bv(\bx))$.
When $\nabla \bv / v \ll \nabla F_a / F_a$, the above is approximately the stress-energy of a perfect fluid of bulk velocity $\bv(\bx)$.
This scenario covers an important class of ``moving fields'' that we wish to generate.

For concreteness, consider a homogeneous Gaussian random field (GRF) $F(t,\bx)$ with zero mean, prescribed by spectra $\Delta_F^2(k)$ and $\Delta_{\partial_t F}^2(k)$ peaked at characteristic wavenumber $k_\ast$.
Realizations of this random field contain $\order{1}$ fluctuations in energy and momentum densities at length scale $\sim k_\ast^{-1}$.
However, the stress-energy appears smooth when coarse grained over a scale larger than $k_\ast^{-1}$, with the smoothed stress-energy being approximately $\expval{T\indices{^\mu_\nu}(\bx)}$.
Upon boosting, the field remains smooth when coarse grained, with stress-energy given by $\expval{\tilde{T}\indices{^\mu_\nu}(\bx)}$.

The criterion $\nabla \bv / v \ll \nabla F_a / F_a$ has a natural interpretation: the velocity perturbations vary on larger scales than the field perturbations.
If $\nabla \bv / v > k_\ast$, the spatially varying boost alters the field at length scales smaller than $k_\ast^{-1}$, which manifests as small scale field fluctuations rather than bulk motion.

The boost will generally distort the field spectrum $\Delta_F^2(k)$, and the condition to avoid this distortion is not $\nabla \bv / v \ll k_\ast$.
To see this intuitively, suppose $F$ is a free field with mass $m$, then the field can be viewed as consisting of wave packets with characteristic speed $v_\ast = k_\ast / \sqrt{k_\ast^2 + m^2}$~\cite{Ling:2024qfv}.
The boost transforms wave packet speeds to $ \sim (v + v_\ast) / (1 + v v_\ast)$, which is approximately $v_\ast$ only if the rapidity $w = \tanh^{-1}(v) \ll \tanh^{-1}(v_\ast)$.
If this condition does not hold, then the characteristic wavenumber after the boost is no longer $k_\ast$, and hence the spectrum $\Delta_F^2(k)$ is no longer peaked at $k_\ast$.

\section{Ultra-relativistic Sine-Gordon soliton}
The Sine-Gordon equation, $\partial_t^2 u - \partial_x^2 u + \sin(u) = 0$, is a nonlinear wave equation that arises in diverse areas of physics, including nonlinear dynamics, condensed matter physics and quantum field theory~\cite{Barone1971,Coleman:1974bu,Mandelstam:1975hb,Callan:1982au,Malomed:2014eba}.
Interestingly, the Sine-Gordon equation admits time-periodic ``breather'' solutions:
\begin{align}
 u_{\mathrm{br}}(t,x) = 4 \tan^{-1} \left( \frac{\sqrt{1-\omega^2}}{\omega} \frac{\sin(\omega t)}{\cosh(\sqrt{1-\omega^2} x)} \right) \,.
\end{align}
This solution is an example of a soliton, whose collision is of both theoretical and phenomenological interest~\cite{Helfer:2018vtq,Amin:2014fua,Eby:2017xaw,Widdicombe:2019woy,Amin:2019ums,Amin:2013eqa,Amin:2013dqa,Kinach:2024hfa,Kinach:2024qzc}.
We demonstrate the spatially varying boost \eqref{eq:infinitesimal_boost} by boosting two initially static breathers to ultra-relativistic speed in opposite directions, thereby setting up initial conditions for subsequent collision.

\begin{figure}[t]
 \centering
  \includegraphics[width=0.5\textwidth]{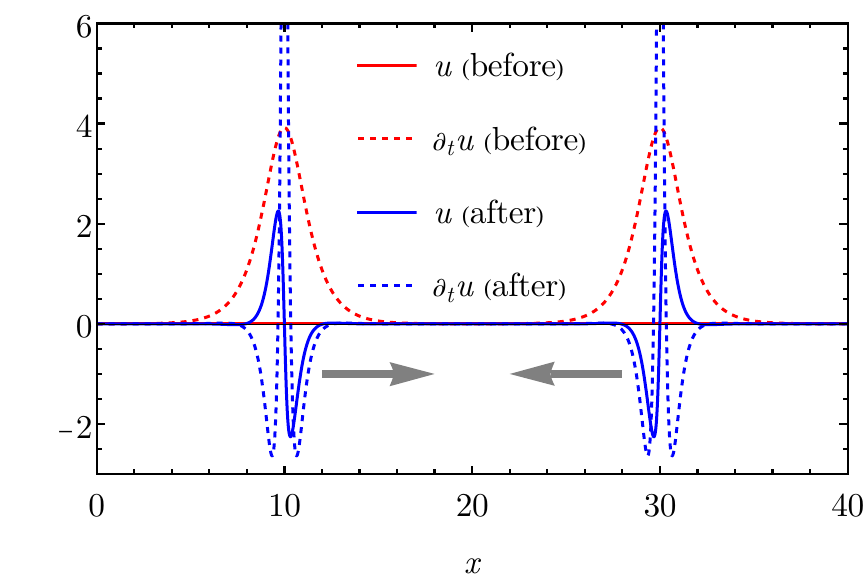}
\caption{The Sine-Gordon initial data before and after boost.
    The two breathers are boosted to opposite velocities $v = \pm 0.964$ by a single spatially varying boost.
  }
  \label{fig:sg_profile_plot}
\end{figure}

We place breathers ($\omega=0.2$) at $x = L / 4$ and $x = 3 L / 4$ ($L = 40$).
The spatially varying boost eq.~\eqref{eq:infinitesimal_boost} is performed by solving this equation from $\epsilon = 0$ to $\epsilon = 1$:
\begin{align}
  \label{eq:sg_boost}
  & \tau(x) = 2 \cos(2 \pi x / L) \frac{L}{2 \pi} \nonumber \\
  & \dv{\epsilon} u(x) = \tau(x) \partial_tu(x) \nonumber \\
  & \dv{\epsilon} \partial_t u(x) = \tau [\partial_x^2 u - \sin(u)]    + (\partial_x\tau) (\partial_x u) \,.
\end{align}
Around $x = L/4$ and $x = 3L/4$, the local rapidity is $w(x) = - \partial_x \tau(x) = \pm 2$, corresponding to velocity $v(x) = \pm \tanh(2) = \pm 0.964$.
Consequently, the boosted breathers move toward each other at velocity $\pm 0.964$.
See fig.~\ref{fig:sg_profile_plot}.

The above example only serves to demonstrate the validity of spatially varying boost; the generated initial data are essentially identical to those obtained by superposing two uniformly boosted breathers.
The benefit of spatially varying boost lies in its use for constructing novel field configurations.
For instance, it enables assignment of large local velocities to a single soliton body, thereby facilitating controlled simulations of soliton deformation or rupture.
Moreover, the method is readily applicable to numerically defined soliton profiles without analytic expressions, such as those dynamically formed in lattice simulations.

\section{Relativistic Proca field}
\label{sec:procedure_proca}

Hypothetical Proca fields (e.g. ``dark photon'') can be produced in the early universe through various mechanisms~\cite{Graham:2015rva, Ahmed:2019mjo,Ema:2019yrd,Ahmed:2020fhc,Kolb:2020fwh,Duch:2017khv,Barman:2020ifq,Barman:2021qds,Agrawal:2017eqm,Agrawal:2018vin,Dror:2018pdh,Co:2018lka,Bastero-Gil:2018uel,Salehian:2020asa,Nelson:2011sf,Arias:2012az,Nakayama:2019rhg,Kitajima:2023fun}.
On sub-horizon scales, these Proca fields can inherit non-relativistic bulk velocities from their progenitors.
Here, we demonstrate the spatially varying boost \eqref{eq:varying_velocity_boosted_ic}  by assigning a relativistic transverse Proca field with large scale velocities $v \ll 1$.
The Proca Lagrangian is $\mathcal{L} = -\frac{1}{4} F_{\mu\nu} F^{\mu\nu} - \frac12 m^2 A_\mu A^\mu$, where $F_{\mu\nu} = \partial_\mu A_\nu - \partial_\nu A_\mu$.

We initialized each field component $A_i$ ($i=1,2,3$) as a homogeneous GRF with white noise spectrum cutoff at $k_\ast = 5 m$.
The field $\bA$ was then projected to the transverse subspace by applying $\delta\indices{^i_j} - k^i k_j / k^2$ in Fourier space.
For the velocity, we set $\tau = \nabla^{-2} f$, where $f$ was generated as a scale invariant GRF with standard deviation $0.1 m$ and cutoff at $k_f = m$.
This choice for $\tau$ translates to $k^3\expval{\bv_\bk \cdot \bv_\bk} \propto k^{-2}$, which emphasizes large scale velocities.

\begin{figure}[t]
  \centering
  \includegraphics[width=0.5\textwidth]{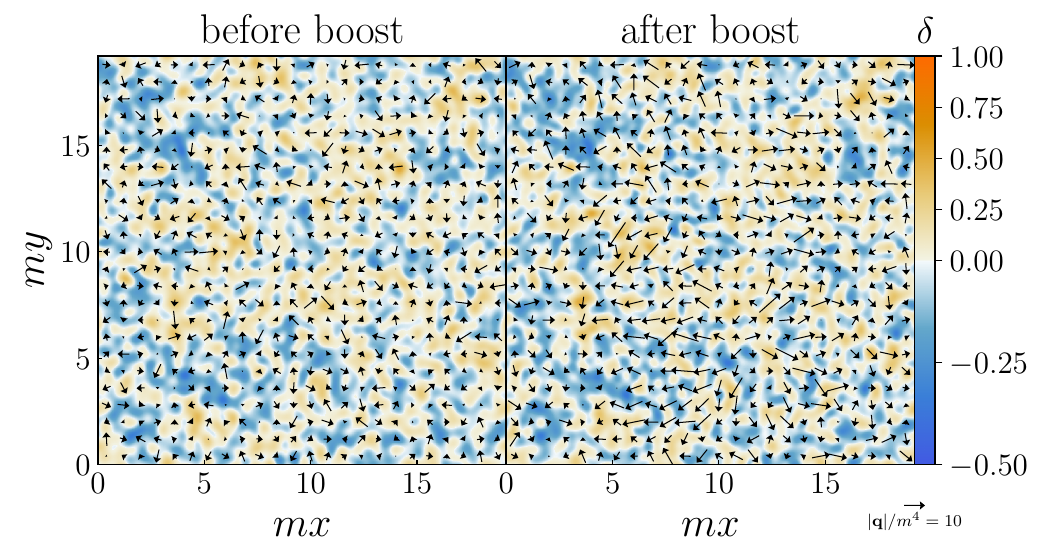}
  \caption{Snapshot of the Proca field overdensity $\delta = \rho / \bar{\rho} - 1$ and momentum density $\bq$ before and after boost.
    Quantities $\delta$ and $\bq$ are averaged over the $z$ axis (perpendicular to the page).
    Large scale velocity perturbations emerge due to the boost, whereas density perturbations appear statistically unaltered.
  }
  \label{fig:proca_snapshot}
\end{figure}

We use the spatially varying boost \eqref{eq:varying_velocity_boosted_ic} for $v \ll 1$:
\begin{align}
  \hat{\bA}(0, \bx) &= \bA - \bv A_t \nonumber \\
  \partial_t \hat{\bA}(0, \bx) &= \partial_t \bA - \bv \partial_t A_t - \bv \cdot \nabla \bA \,,
\end{align}
where the RHS is evaluated at $(\tau(\bx),\bx)$.
Fig.~\ref{fig:proca_snapshot} visualizes the Proca field before and after the boost.
Before boost, the density and velocity perturbation are dominated by fluctuations on length scale $2 \pi / k_\ast \sim 1$, and the field is statistically homogeneous.
After the boost, there are clearly velocity perturbations on scales larger than $k_\ast^{-1}$, but the density perturbations are mostly undisturbed.

\section{Spin-$1$ wave dark matter}
\label{sec:procedure_non-relativistic_vector}
Wave dark matter is a dark matter candidate with rich phenomenology~\cite{Hui:2016ltb,Hu:2000ke,Ferreira:2020fam,Hui:2021tkt}.
Being wave like, its time evolution is treated with the \Schr-Poisson (SP) formalism, an effective field theory for non-relativistic fields~\cite{Salehian:2021khb,Salehian:2020bon,Namjoo:2017nia,Braaten:2018lmj}.
SP simulations are thus useful for studying various aspects of wave dark matter phenomenology, including structure formation, large and small scale structure, halo and subhalo mass function, solitonic configurations, kinetic relaxation and free streaming~\cite{Schive:2014dra,Schive:2014hza,Veltmaat:2016rxo,Edwards:2018ccc,Li:2018kyk,Veltmaat:2018dfz,Bar:2018acw,Li:2020ryg,Chen:2020cef,May:2021wwp,Chan:2021bja,Jain:2022agt,May:2022gus,Amin:2022pzv,Jain:2023tsr,Ling:2024qfv}.
Here, we apply spatially varying boost to generate a spin-$1$ wave dark matter field with characteristic wavenumber $k_\ast$, and with consistent density and velocity perturbations during matter domination.

For a Proca field $\bA$ in a Friedmann-Robertson-Walker (FRW) spacetime, the corresponding \Schr-Poisson field $\bPsi = (\psi_1, \psi_2, \psi_3)$ is defined by the ansatz
\begin{align}
  \label{eq:schr_poisson_ansatz}
  \bA(t, \bx) = \frac{1}{\sqrt{2 m a(t)}} \left[ \bPsi(t,\bx) e^{-imt} + \text{c.c.} \right] \,,
\end{align}
where $a(t)$ is the scale factor.
Its energy and momentum densities are~\cite{Nambo:2024hao,Jain:2022agt,Amin:2022pzv}
\begin{align}
  \rho &= a^{-3} (1 - \Phi) m \bPsi^\dagger \cdot \bPsi,\quad
  \bq = a^{-4} \Re \left[i \psi_j \nabla \psi_j^\dagger\right] \,,
\end{align}
where $\Phi$ is the Newtonian gravitational potential.

For non-relativistic $\bA$ and $v \ll 1$, the spatial gradient terms in eq.~\eqref{eq:varying_velocity_boosted_ic} are all negligible, and $A_i$ has approximate solution $A_i(t,\bx) \approx c_1(\bx) \cos(mt) + c_2(\bx) \sin(mt)$ for short time periods $t \ll t_{\mathrm{deBroglie}} = (k_*^2 / m)^{-1}$.
Combining this solution with \eqref{eq:schr_poisson_ansatz}, the spatially varying boost \eqref{eq:varying_velocity_boosted_ic} becomes a simple phase shift:
\begin{align}
  \label{eq:varying_galilean_boost}
  \hat{\bPsi}(0,\bx) \approx \bPsi(\tau(\bx),\bx) \approx e^{-i m \tau(\bx)} \bPsi(0,\bx) \,.
\end{align}
This transform can be understood as a spatially varying Galilean boost.
One can verify that $\rho(\bx) \to \rho(\bx)$ and $\bq(\bx) \to \bq(\bx) + \rho(\bx) \bv(\bx)$ under this boost.
Note that there is no need for time-stepping to implement eq.~\eqref{eq:varying_galilean_boost} numerically.

Cosmological perturbation theory predicts the large scale density and velocity perturbations for dark matter~\cite{Hu:1995en,Baumann:2022mni}.
On sub-horizon length scales during matter domination, the density perturbation $\delta = \rho / \bar{\rho} - 1$ is given in Fourier space by
\begin{align}
  \label{eq:cpt_prediction}
  \delta_{\bk}(t) &= - \left(\frac{a_{\mathrm{eq}}}{a_i}\right)^2 \frac{a(t)}{a_{\mathrm{eq}}} \mathcal{R}_{\bk} \times
  \begin{cases}
    1 & k < k_{\mathrm{eq}} \\
    \left(\frac{a_k}{a_{\mathrm{eq}}}\right)^2 & k > k_{\mathrm{eq}}
  \end{cases}
  \nonumber \\
  \Phi_\bk(t) &= \frac35 \mathcal{R}_\bk \times \Theta(k_{\mathrm{eq}} - k) 
                \end{align}
in the notation of ref.~\cite{Baumann:2022mni}.
The velocity is thus fixed by the continuity equation $\partial_t \delta_{\bk} + \nabla \cdot \bv / a - 3 \partial_t \Phi = 0$.

\begin{figure}[t]
  \centering
  \includegraphics[width=0.5\textwidth]{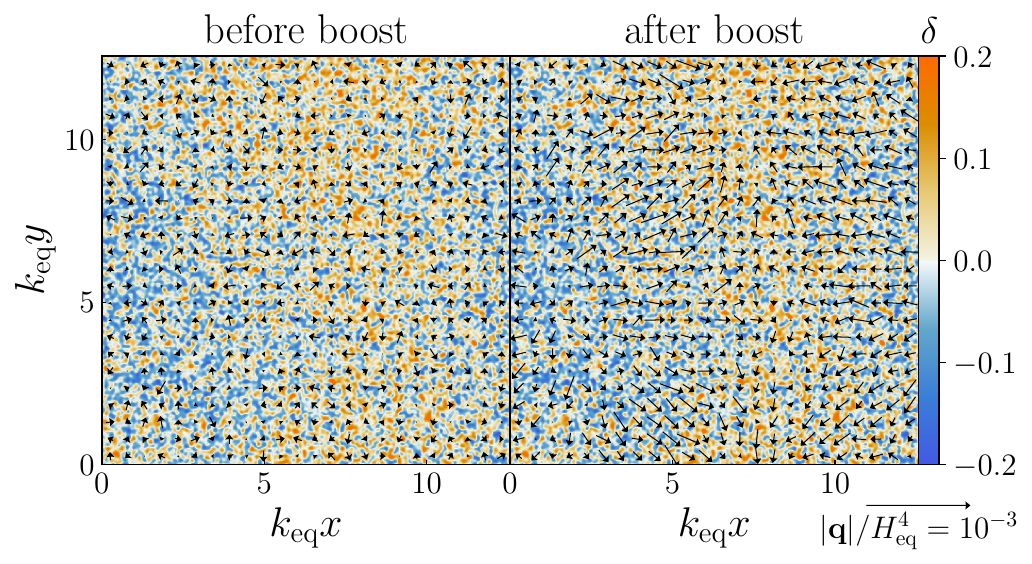}
  \caption{Snapshot of energy overdensity and momentum density of $\bPsi$ before and after boost.
    Before boost, velocities pointing in random directions due to small scale field fluctuations.
    After boost, bulk velocities are pointing in the high overdensity regions, visualizing the bulk motion that lead to density build up.
  }
  \label{fig:sp_snapshot}
\end{figure}

In accordance with cosmological perturbation theory, we generated a GRF $\mathcal{R}_{\bk}$ with scale invariant spectrum.
We then used the inhomogeneous GRF algorithm described in ref.~\cite{Ling:2024qfv} to generate $\psi_i$'s with white noise spectrum cutoff at $k_\ast$, and with variance perturbation $\expval{|\psi_i|^2(\bx)} = (1 + \delta(\bx) + \Phi(\bx)) \overline{|\psi_i|^2}$.
Finally, we boosted $\bPsi$ with eq.~\eqref{eq:varying_galilean_boost} given $\tau = a^2 \nabla^{-2}[\partial_t \delta - 3 \partial_t \Phi]$.
The boosted field $\hat{\bPsi}$ has density and velocity perturbations matching eq.~\eqref{eq:cpt_prediction}.
See fig.~\ref{fig:sp_snapshot} for an illustration.

While previous works already initialize \Schr-Poisson fields with appropriate velocities~\cite{Veltmaat:2018dfz,May:2022gus}, their methods do not allow for a separate scale $k_\ast$ of small scale fluctuations.
As discussed in refs.~\cite{Amin:2022nlh,Ling:2024qfv,Liu:2024pjg,Amin:2025dtd,Liu:2025lts,Capanelli:2025nrj,Amin:2025sla}, the field's wavenumber $k_\ast$ can be much higher than $\nabla \bv / v$, and its value is phenomenologically important due to effects such as free streaming and wave interference.
Our scheme (which was recently employed in ref.~\cite{Amin:2025sla}) allows for setting $k_\ast$ to be a value of choice, which enables controlled numerical experiments for halo or structure formation that incorporate the corresponding physics.

\section{Discussion}
\label{sec:discussion}
The spatially varying boost algorithm provides a powerful method for endowing field initial data with prescribed velocities, including ultra-relativistic motion.
Crucially, the approach preserves the local character of the field, separating generation of small scale behavior from bulk velocity.
This capability enables novel first-principles simulations across diverse physics disciplines, simultaneously capturing both microscopic field dynamics and emergent bulk behavior.
For phenomena where bulk velocities play a critical role, our algorithm now facilitates controlled numerical experiments that were previously inaccessible.
We have made the code for this work available \href{https://github.com/hypermania/Cosmic-Fields-Lite}{here} .

Finally, we comment on future directions of research.
First, that the algorithm encodes velocity (rapidity) via $\tau(x)$ with $\bv(x) = - \nabla \tau(x)$ restricts the velocity (rapidity) fields to be irrotational.
One line of research is to remove this limitation.
Second, the boost eq.~\eqref{eq:infinitesimal_boost} is nonlinear for nonlinear field theories, so it would create non-Gaussianities from Gaussian initial data.
One avenue is to explore whether this algorithm can be used to generate spatially varying non-Gaussianities, thereby extending literature works on generating such phenomenologically interesting initial conditions~\cite{Scoccimarro:2011pz,Coulton:2023oug,Bao:2025onc}.

\section{Acknowledgments}

\begin{acknowledgments}
  I thank Mustafa A. Amin, Kaden Hazzard, Andrew J. Long, Han Pu and Yiming Zhong for useful discussions and critical comments.
  I extend special thanks to Mustafa A. Amin for his early involvement in this paper and discussions on cosmological perturbations for wave dark matter.
  I also thank Yuxuan He, Sida Lu, Yue Nan, Andrey Shkerin and Sam S. C. Wong for stimulating discussions.
  This project is supported by APRC-CityU New Research Initiatives/Infrastructure Support from Central.
\end{acknowledgments}

\bibliography{main,manual}\newpage

\begin{thebibliography}{117}%
\makeatletter
\providecommand \@ifxundefined [1]{%
 \@ifx{#1\undefined}
}%
\providecommand \@ifnum [1]{%
 \ifnum #1\expandafter \@firstoftwo
 \else \expandafter \@secondoftwo
 \fi
}%
\providecommand \@ifx [1]{%
 \ifx #1\expandafter \@firstoftwo
 \else \expandafter \@secondoftwo
 \fi
}%
\providecommand \natexlab [1]{#1}%
\providecommand \enquote  [1]{``#1''}%
\providecommand \bibnamefont  [1]{#1}%
\providecommand \bibfnamefont [1]{#1}%
\providecommand \citenamefont [1]{#1}%
\providecommand \href@noop [0]{\@secondoftwo}%
\providecommand \href [0]{\begingroup \@sanitize@url \@href}%
\providecommand \@href[1]{\@@startlink{#1}\@@href}%
\providecommand \@@href[1]{\endgroup#1\@@endlink}%
\providecommand \@sanitize@url [0]{\catcode `\\12\catcode `\$12\catcode
  `\&12\catcode `\#12\catcode `\^12\catcode `\_12\catcode `\%12\relax}%
\providecommand \@@startlink[1]{}%
\providecommand \@@endlink[0]{}%
\providecommand \url  [0]{\begingroup\@sanitize@url \@url }%
\providecommand \@url [1]{\endgroup\@href {#1}{\urlprefix }}%
\providecommand \urlprefix  [0]{URL }%
\providecommand \Eprint [0]{\href }%
\providecommand \doibase [0]{https://doi.org/}%
\providecommand \selectlanguage [0]{\@gobble}%
\providecommand \bibinfo  [0]{\@secondoftwo}%
\providecommand \bibfield  [0]{\@secondoftwo}%
\providecommand \translation [1]{[#1]}%
\providecommand \BibitemOpen [0]{}%
\providecommand \bibitemStop [0]{}%
\providecommand \bibitemNoStop [0]{.\EOS\space}%
\providecommand \EOS [0]{\spacefactor3000\relax}%
\providecommand \BibitemShut  [1]{\csname bibitem#1\endcsname}%
\let\auto@bib@innerbib\@empty
%</preamble>
\bibitem [{\citenamefont {Efstathiou}\ \emph {et~al.}(1985)\citenamefont
  {Efstathiou}, \citenamefont {Davis}, \citenamefont {Frenk},\ and\
  \citenamefont {White}}]{Efstathiou:1985re}%
  \BibitemOpen
  \bibfield  {author} {\bibinfo {author} {\bibfnamefont {G.}~\bibnamefont
  {Efstathiou}}, \bibinfo {author} {\bibfnamefont {M.}~\bibnamefont {Davis}},
  \bibinfo {author} {\bibfnamefont {C.~S.}\ \bibnamefont {Frenk}},\ and\
  \bibinfo {author} {\bibfnamefont {S.~D.~M.}\ \bibnamefont {White}},\
  }\bibfield  {title} {\bibinfo {title} {{Numerical Techniques for Large
  Cosmological N-Body Simulations}},\ }\href {https://doi.org/10.1086/191003}
  {\bibfield  {journal} {\bibinfo  {journal} {Astrophys. J. Suppl.}\ }\textbf
  {\bibinfo {volume} {57}},\ \bibinfo {pages} {241} (\bibinfo {year}
  {1985})}\BibitemShut {NoStop}%
\bibitem [{\citenamefont {Matarrese}\ \emph {et~al.}(1993)\citenamefont
  {Matarrese}, \citenamefont {Pantano},\ and\ \citenamefont
  {Saez}}]{Matarrese:1992rp}%
  \BibitemOpen
  \bibfield  {author} {\bibinfo {author} {\bibfnamefont {S.}~\bibnamefont
  {Matarrese}}, \bibinfo {author} {\bibfnamefont {O.}~\bibnamefont {Pantano}},\
  and\ \bibinfo {author} {\bibfnamefont {D.}~\bibnamefont {Saez}},\ }\bibfield
  {title} {\bibinfo {title} {{A General relativistic approach to the nonlinear
  evolution of collisionless matter}},\ }\href
  {https://doi.org/10.1103/PhysRevD.47.1311} {\bibfield  {journal} {\bibinfo
  {journal} {Phys. Rev. D}\ }\textbf {\bibinfo {volume} {47}},\ \bibinfo
  {pages} {1311} (\bibinfo {year} {1993})}\BibitemShut {NoStop}%
\bibitem [{\citenamefont {Scoccimarro}(1998)}]{Scoccimarro:1997gr}%
  \BibitemOpen
  \bibfield  {author} {\bibinfo {author} {\bibfnamefont {R.}~\bibnamefont
  {Scoccimarro}},\ }\bibfield  {title} {\bibinfo {title} {{Transients from
  initial conditions: a perturbative analysis}},\ }\href
  {https://doi.org/10.1046/j.1365-8711.1998.01845.x} {\bibfield  {journal}
  {\bibinfo  {journal} {Mon. Not. Roy. Astron. Soc.}\ }\textbf {\bibinfo
  {volume} {299}},\ \bibinfo {pages} {1097} (\bibinfo {year} {1998})},\ \Eprint
  {https://arxiv.org/abs/astro-ph/9711187} {arXiv:astro-ph/9711187}
  \BibitemShut {NoStop}%
\bibitem [{\citenamefont {Valageas}(2002)}]{Valageas:2001qe}%
  \BibitemOpen
  \bibfield  {author} {\bibinfo {author} {\bibfnamefont {P.}~\bibnamefont
  {Valageas}},\ }\bibfield  {title} {\bibinfo {title} {{Transients from
  zel'dovich initial conditions}},\ }\href
  {https://doi.org/10.1051/0004-6361:20020187} {\bibfield  {journal} {\bibinfo
  {journal} {Astron. Astrophys.}\ }\textbf {\bibinfo {volume} {385}},\ \bibinfo
  {pages} {761} (\bibinfo {year} {2002})},\ \Eprint
  {https://arxiv.org/abs/astro-ph/0112102} {arXiv:astro-ph/0112102}
  \BibitemShut {NoStop}%
\bibitem [{\citenamefont {Yoshida}\ \emph {et~al.}(2003)\citenamefont
  {Yoshida}, \citenamefont {Sugiyama},\ and\ \citenamefont
  {Hernquist}}]{Yoshida:2003sy}%
  \BibitemOpen
  \bibfield  {author} {\bibinfo {author} {\bibfnamefont {N.}~\bibnamefont
  {Yoshida}}, \bibinfo {author} {\bibfnamefont {N.}~\bibnamefont {Sugiyama}},\
  and\ \bibinfo {author} {\bibfnamefont {L.}~\bibnamefont {Hernquist}},\
  }\bibfield  {title} {\bibinfo {title} {{The evolution of baryon density
  fluctuations in multi-component cosmological simulations}},\ }\href
  {https://doi.org/10.1046/j.1365-8711.2003.06829.x} {\bibfield  {journal}
  {\bibinfo  {journal} {Mon. Not. Roy. Astron. Soc.}\ }\textbf {\bibinfo
  {volume} {344}},\ \bibinfo {pages} {481} (\bibinfo {year} {2003})},\ \Eprint
  {https://arxiv.org/abs/astro-ph/0305210} {arXiv:astro-ph/0305210}
  \BibitemShut {NoStop}%
\bibitem [{\citenamefont {Sirko}(2005)}]{Sirko:2005uz}%
  \BibitemOpen
  \bibfield  {author} {\bibinfo {author} {\bibfnamefont {E.}~\bibnamefont
  {Sirko}},\ }\bibfield  {title} {\bibinfo {title} {{Initial conditions to
  cosmological N-body simulations, or how to run an ensemble of simulations}},\
  }\href {https://doi.org/10.1086/497090} {\bibfield  {journal} {\bibinfo
  {journal} {Astrophys. J.}\ }\textbf {\bibinfo {volume} {634}},\ \bibinfo
  {pages} {728} (\bibinfo {year} {2005})},\ \Eprint
  {https://arxiv.org/abs/astro-ph/0503106} {arXiv:astro-ph/0503106}
  \BibitemShut {NoStop}%
\bibitem [{\citenamefont {Crocce}\ \emph {et~al.}(2006)\citenamefont {Crocce},
  \citenamefont {Pueblas},\ and\ \citenamefont {Scoccimarro}}]{Crocce:2006ve}%
  \BibitemOpen
  \bibfield  {author} {\bibinfo {author} {\bibfnamefont {M.}~\bibnamefont
  {Crocce}}, \bibinfo {author} {\bibfnamefont {S.}~\bibnamefont {Pueblas}},\
  and\ \bibinfo {author} {\bibfnamefont {R.}~\bibnamefont {Scoccimarro}},\
  }\bibfield  {title} {\bibinfo {title} {{Transients from Initial Conditions in
  Cosmological Simulations}},\ }\href
  {https://doi.org/10.1111/j.1365-2966.2006.11040.x} {\bibfield  {journal}
  {\bibinfo  {journal} {Mon. Not. Roy. Astron. Soc.}\ }\textbf {\bibinfo
  {volume} {373}},\ \bibinfo {pages} {369} (\bibinfo {year} {2006})},\ \Eprint
  {https://arxiv.org/abs/astro-ph/0606505} {arXiv:astro-ph/0606505}
  \BibitemShut {NoStop}%
\bibitem [{\citenamefont {Tatekawa}\ and\ \citenamefont
  {Mizuno}(2007)}]{Tatekawa:2007ix}%
  \BibitemOpen
  \bibfield  {author} {\bibinfo {author} {\bibfnamefont {T.}~\bibnamefont
  {Tatekawa}}\ and\ \bibinfo {author} {\bibfnamefont {S.}~\bibnamefont
  {Mizuno}},\ }\bibfield  {title} {\bibinfo {title} {{Transients from initial
  conditions based on Lagrangian perturbation theory in $N$-body
  simulations}},\ }\href {https://doi.org/10.1088/1475-7516/2007/12/014}
  {\bibfield  {journal} {\bibinfo  {journal} {JCAP}\ }\textbf {\bibinfo
  {volume} {12}},\ \bibinfo {pages} {014}},\ \Eprint
  {https://arxiv.org/abs/0706.1334} {arXiv:0706.1334 [astro-ph]} \BibitemShut
  {NoStop}%
\bibitem [{\citenamefont {Hahn}\ and\ \citenamefont
  {Abel}(2011)}]{Hahn:2011uy}%
  \BibitemOpen
  \bibfield  {author} {\bibinfo {author} {\bibfnamefont {O.}~\bibnamefont
  {Hahn}}\ and\ \bibinfo {author} {\bibfnamefont {T.}~\bibnamefont {Abel}},\
  }\bibfield  {title} {\bibinfo {title} {{Multi-scale initial conditions for
  cosmological simulations}},\ }\href
  {https://doi.org/10.1111/j.1365-2966.2011.18820.x} {\bibfield  {journal}
  {\bibinfo  {journal} {Mon. Not. Roy. Astron. Soc.}\ }\textbf {\bibinfo
  {volume} {415}},\ \bibinfo {pages} {2101} (\bibinfo {year} {2011})},\ \Eprint
  {https://arxiv.org/abs/1103.6031} {arXiv:1103.6031 [astro-ph.CO]}
  \BibitemShut {NoStop}%
\bibitem [{\citenamefont {L'Huillier}\ \emph {et~al.}(2014)\citenamefont
  {L'Huillier}, \citenamefont {Park},\ and\ \citenamefont
  {Kim}}]{LHuillier:2014fcj}%
  \BibitemOpen
  \bibfield  {author} {\bibinfo {author} {\bibfnamefont {B.}~\bibnamefont
  {L'Huillier}}, \bibinfo {author} {\bibfnamefont {C.}~\bibnamefont {Park}},\
  and\ \bibinfo {author} {\bibfnamefont {J.}~\bibnamefont {Kim}},\ }\bibfield
  {title} {\bibinfo {title} {{Effects of the initial conditions on cosmological
  $N$-body simulations}},\ }\href
  {https://doi.org/10.1016/j.newast.2014.01.007} {\bibfield  {journal}
  {\bibinfo  {journal} {New Astron.}\ }\textbf {\bibinfo {volume} {30}},\
  \bibinfo {pages} {79} (\bibinfo {year} {2014})},\ \Eprint
  {https://arxiv.org/abs/1401.6180} {arXiv:1401.6180 [astro-ph.CO]}
  \BibitemShut {NoStop}%
\bibitem [{\citenamefont {Hahn}\ \emph {et~al.}(2015)\citenamefont {Hahn},
  \citenamefont {Angulo},\ and\ \citenamefont {Abel}}]{Hahn:2014lca}%
  \BibitemOpen
  \bibfield  {author} {\bibinfo {author} {\bibfnamefont {O.}~\bibnamefont
  {Hahn}}, \bibinfo {author} {\bibfnamefont {R.~E.}\ \bibnamefont {Angulo}},\
  and\ \bibinfo {author} {\bibfnamefont {T.}~\bibnamefont {Abel}},\ }\bibfield
  {title} {\bibinfo {title} {{The Properties of Cosmic Velocity Fields}},\
  }\href {https://doi.org/10.1093/mnras/stv2179} {\bibfield  {journal}
  {\bibinfo  {journal} {Mon. Not. Roy. Astron. Soc.}\ }\textbf {\bibinfo
  {volume} {454}},\ \bibinfo {pages} {3920} (\bibinfo {year} {2015})},\ \Eprint
  {https://arxiv.org/abs/1404.2280} {arXiv:1404.2280 [astro-ph.CO]}
  \BibitemShut {NoStop}%
\bibitem [{\citenamefont {Fidler}\ \emph {et~al.}(2015)\citenamefont {Fidler},
  \citenamefont {Rampf}, \citenamefont {Tram}, \citenamefont {Crittenden},
  \citenamefont {Koyama},\ and\ \citenamefont {Wands}}]{Fidler:2015npa}%
  \BibitemOpen
  \bibfield  {author} {\bibinfo {author} {\bibfnamefont {C.}~\bibnamefont
  {Fidler}}, \bibinfo {author} {\bibfnamefont {C.}~\bibnamefont {Rampf}},
  \bibinfo {author} {\bibfnamefont {T.}~\bibnamefont {Tram}}, \bibinfo {author}
  {\bibfnamefont {R.}~\bibnamefont {Crittenden}}, \bibinfo {author}
  {\bibfnamefont {K.}~\bibnamefont {Koyama}},\ and\ \bibinfo {author}
  {\bibfnamefont {D.}~\bibnamefont {Wands}},\ }\bibfield  {title} {\bibinfo
  {title} {{General relativistic corrections to $N$-body simulations and the
  Zel'dovich approximation}},\ }\href
  {https://doi.org/10.1103/PhysRevD.92.123517} {\bibfield  {journal} {\bibinfo
  {journal} {Phys. Rev. D}\ }\textbf {\bibinfo {volume} {92}},\ \bibinfo
  {pages} {123517} (\bibinfo {year} {2015})},\ \Eprint
  {https://arxiv.org/abs/1505.04756} {arXiv:1505.04756 [astro-ph.CO]}
  \BibitemShut {NoStop}%
\bibitem [{\citenamefont {Adamek}\ \emph
  {et~al.}(2016{\natexlab{a}})\citenamefont {Adamek}, \citenamefont {Daverio},
  \citenamefont {Durrer},\ and\ \citenamefont {Kunz}}]{Adamek:2015eda}%
  \BibitemOpen
  \bibfield  {author} {\bibinfo {author} {\bibfnamefont {J.}~\bibnamefont
  {Adamek}}, \bibinfo {author} {\bibfnamefont {D.}~\bibnamefont {Daverio}},
  \bibinfo {author} {\bibfnamefont {R.}~\bibnamefont {Durrer}},\ and\ \bibinfo
  {author} {\bibfnamefont {M.}~\bibnamefont {Kunz}},\ }\bibfield  {title}
  {\bibinfo {title} {{General relativity and cosmic structure formation}},\
  }\href {https://doi.org/10.1038/nphys3673} {\bibfield  {journal} {\bibinfo
  {journal} {Nature Phys.}\ }\textbf {\bibinfo {volume} {12}},\ \bibinfo
  {pages} {346} (\bibinfo {year} {2016}{\natexlab{a}})},\ \Eprint
  {https://arxiv.org/abs/1509.01699} {arXiv:1509.01699 [astro-ph.CO]}
  \BibitemShut {NoStop}%
\bibitem [{\citenamefont {Adamek}\ \emph
  {et~al.}(2016{\natexlab{b}})\citenamefont {Adamek}, \citenamefont {Daverio},
  \citenamefont {Durrer},\ and\ \citenamefont {Kunz}}]{Adamek:2016zes}%
  \BibitemOpen
  \bibfield  {author} {\bibinfo {author} {\bibfnamefont {J.}~\bibnamefont
  {Adamek}}, \bibinfo {author} {\bibfnamefont {D.}~\bibnamefont {Daverio}},
  \bibinfo {author} {\bibfnamefont {R.}~\bibnamefont {Durrer}},\ and\ \bibinfo
  {author} {\bibfnamefont {M.}~\bibnamefont {Kunz}},\ }\bibfield  {title}
  {\bibinfo {title} {{gevolution: a cosmological N-body code based on General
  Relativity}},\ }\href {https://doi.org/10.1088/1475-7516/2016/07/053}
  {\bibfield  {journal} {\bibinfo  {journal} {JCAP}\ }\textbf {\bibinfo
  {volume} {07}},\ \bibinfo {pages} {053}},\ \Eprint
  {https://arxiv.org/abs/1604.06065} {arXiv:1604.06065 [astro-ph.CO]}
  \BibitemShut {NoStop}%
\bibitem [{\citenamefont {Zennaro}\ \emph {et~al.}(2017)\citenamefont
  {Zennaro}, \citenamefont {Bel}, \citenamefont {Villaescusa-Navarro},
  \citenamefont {Carbone}, \citenamefont {Sefusatti},\ and\ \citenamefont
  {Guzzo}}]{Zennaro:2016nqo}%
  \BibitemOpen
  \bibfield  {author} {\bibinfo {author} {\bibfnamefont {M.}~\bibnamefont
  {Zennaro}}, \bibinfo {author} {\bibfnamefont {J.}~\bibnamefont {Bel}},
  \bibinfo {author} {\bibfnamefont {F.}~\bibnamefont {Villaescusa-Navarro}},
  \bibinfo {author} {\bibfnamefont {C.}~\bibnamefont {Carbone}}, \bibinfo
  {author} {\bibfnamefont {E.}~\bibnamefont {Sefusatti}},\ and\ \bibinfo
  {author} {\bibfnamefont {L.}~\bibnamefont {Guzzo}},\ }\bibfield  {title}
  {\bibinfo {title} {{Initial Conditions for Accurate N-Body Simulations of
  Massive Neutrino Cosmologies}},\ }\href
  {https://doi.org/10.1093/mnras/stw3340} {\bibfield  {journal} {\bibinfo
  {journal} {Mon. Not. Roy. Astron. Soc.}\ }\textbf {\bibinfo {volume} {466}},\
  \bibinfo {pages} {3244} (\bibinfo {year} {2017})},\ \Eprint
  {https://arxiv.org/abs/1605.05283} {arXiv:1605.05283 [astro-ph.CO]}
  \BibitemShut {NoStop}%
\bibitem [{\citenamefont {Garrison}\ \emph {et~al.}(2016)\citenamefont
  {Garrison}, \citenamefont {Eisenstein}, \citenamefont {Ferrer}, \citenamefont
  {Metchnik},\ and\ \citenamefont {Pinto}}]{Garrison:2016vvp}%
  \BibitemOpen
  \bibfield  {author} {\bibinfo {author} {\bibfnamefont {L.~H.}\ \bibnamefont
  {Garrison}}, \bibinfo {author} {\bibfnamefont {D.~J.}\ \bibnamefont
  {Eisenstein}}, \bibinfo {author} {\bibfnamefont {D.}~\bibnamefont {Ferrer}},
  \bibinfo {author} {\bibfnamefont {M.~V.}\ \bibnamefont {Metchnik}},\ and\
  \bibinfo {author} {\bibfnamefont {P.~A.}\ \bibnamefont {Pinto}},\ }\bibfield
  {title} {\bibinfo {title} {{Improving Initial Conditions for Cosmological
  $N$-Body Simulations}},\ }\href {https://doi.org/10.1093/mnras/stw1594}
  {\bibfield  {journal} {\bibinfo  {journal} {Mon. Not. Roy. Astron. Soc.}\
  }\textbf {\bibinfo {volume} {461}},\ \bibinfo {pages} {4125} (\bibinfo {year}
  {2016})},\ \Eprint {https://arxiv.org/abs/1605.02333} {arXiv:1605.02333
  [astro-ph.CO]} \BibitemShut {NoStop}%
\bibitem [{\citenamefont {Tatekawa}(2019)}]{Tatekawa:2019zpc}%
  \BibitemOpen
  \bibfield  {author} {\bibinfo {author} {\bibfnamefont {T.}~\bibnamefont
  {Tatekawa}},\ }\bibfield  {title} {\bibinfo {title} {{Transients from Initial
  Conditions Based on Lagrangian Perturbation Theory in $N$-body Simulations
  III: The Case of GADGET-2 Code}}\ }\href
  {https://doi.org/10.1142/S0218271820500960} {10.1142/S0218271820500960}
  (\bibinfo {year} {2019}),\ \Eprint {https://arxiv.org/abs/1901.09210}
  {arXiv:1901.09210 [astro-ph.CO]} \BibitemShut {NoStop}%
\bibitem [{\citenamefont {Springel}\ \emph {et~al.}(2021)\citenamefont
  {Springel}, \citenamefont {Pakmor}, \citenamefont {Zier},\ and\ \citenamefont
  {Reinecke}}]{Springel:2020plp}%
  \BibitemOpen
  \bibfield  {author} {\bibinfo {author} {\bibfnamefont {V.}~\bibnamefont
  {Springel}}, \bibinfo {author} {\bibfnamefont {R.}~\bibnamefont {Pakmor}},
  \bibinfo {author} {\bibfnamefont {O.}~\bibnamefont {Zier}},\ and\ \bibinfo
  {author} {\bibfnamefont {M.}~\bibnamefont {Reinecke}},\ }\bibfield  {title}
  {\bibinfo {title} {{Simulating cosmic structure formation with the gadget-4
  code}},\ }\href {https://doi.org/10.1093/mnras/stab1855} {\bibfield
  {journal} {\bibinfo  {journal} {Mon. Not. Roy. Astron. Soc.}\ }\textbf
  {\bibinfo {volume} {506}},\ \bibinfo {pages} {2871} (\bibinfo {year}
  {2021})},\ \Eprint {https://arxiv.org/abs/2010.03567} {arXiv:2010.03567
  [astro-ph.IM]} \BibitemShut {NoStop}%
\bibitem [{\citenamefont {Hahn}\ \emph {et~al.}(2021)\citenamefont {Hahn},
  \citenamefont {Rampf},\ and\ \citenamefont {Uhlemann}}]{Hahn:2020lvr}%
  \BibitemOpen
  \bibfield  {author} {\bibinfo {author} {\bibfnamefont {O.}~\bibnamefont
  {Hahn}}, \bibinfo {author} {\bibfnamefont {C.}~\bibnamefont {Rampf}},\ and\
  \bibinfo {author} {\bibfnamefont {C.}~\bibnamefont {Uhlemann}},\ }\bibfield
  {title} {\bibinfo {title} {{Higher order initial conditions for mixed
  baryon{\textendash}CDM simulations}},\ }\href
  {https://doi.org/10.1093/mnras/staa3773} {\bibfield  {journal} {\bibinfo
  {journal} {Mon. Not. Roy. Astron. Soc.}\ }\textbf {\bibinfo {volume} {503}},\
  \bibinfo {pages} {426} (\bibinfo {year} {2021})},\ \Eprint
  {https://arxiv.org/abs/2008.09124} {arXiv:2008.09124 [astro-ph.CO]}
  \BibitemShut {NoStop}%
\bibitem [{\citenamefont {Michaux}\ \emph {et~al.}(2020)\citenamefont
  {Michaux}, \citenamefont {Hahn}, \citenamefont {Rampf},\ and\ \citenamefont
  {Angulo}}]{Michaux:2020yis}%
  \BibitemOpen
  \bibfield  {author} {\bibinfo {author} {\bibfnamefont {M.}~\bibnamefont
  {Michaux}}, \bibinfo {author} {\bibfnamefont {O.}~\bibnamefont {Hahn}},
  \bibinfo {author} {\bibfnamefont {C.}~\bibnamefont {Rampf}},\ and\ \bibinfo
  {author} {\bibfnamefont {R.~E.}\ \bibnamefont {Angulo}},\ }\bibfield  {title}
  {\bibinfo {title} {{Accurate initial conditions for cosmological N-body
  simulations: Minimizing truncation and discreteness errors}},\ }\href
  {https://doi.org/10.1093/mnras/staa3149} {\bibfield  {journal} {\bibinfo
  {journal} {Mon. Not. Roy. Astron. Soc.}\ }\textbf {\bibinfo {volume} {500}},\
  \bibinfo {pages} {663} (\bibinfo {year} {2020})},\ \Eprint
  {https://arxiv.org/abs/2008.09588} {arXiv:2008.09588 [astro-ph.CO]}
  \BibitemShut {NoStop}%
\bibitem [{\citenamefont {Barrera-Hinojosa}\ and\ \citenamefont
  {Li}(2020)}]{Barrera-Hinojosa:2020arz}%
  \BibitemOpen
  \bibfield  {author} {\bibinfo {author} {\bibfnamefont {C.}~\bibnamefont
  {Barrera-Hinojosa}}\ and\ \bibinfo {author} {\bibfnamefont {B.}~\bibnamefont
  {Li}},\ }\bibfield  {title} {\bibinfo {title} {{GRAMSES: a new route to
  general relativistic $N$-body simulations in cosmology. Part II. Initial
  conditions}},\ }\href {https://doi.org/10.1088/1475-7516/2020/04/056}
  {\bibfield  {journal} {\bibinfo  {journal} {JCAP}\ }\textbf {\bibinfo
  {volume} {04}},\ \bibinfo {pages} {056}},\ \Eprint
  {https://arxiv.org/abs/2001.07968} {arXiv:2001.07968 [astro-ph.CO]}
  \BibitemShut {NoStop}%
\bibitem [{\citenamefont {Angulo}\ and\ \citenamefont
  {Hahn}(2022)}]{Angulo:2021kes}%
  \BibitemOpen
  \bibfield  {author} {\bibinfo {author} {\bibfnamefont {R.~E.}\ \bibnamefont
  {Angulo}}\ and\ \bibinfo {author} {\bibfnamefont {O.}~\bibnamefont {Hahn}},\
  }\bibfield  {title} {\bibinfo {title} {{Large-scale dark matter
  simulations}},\ }\href {https://doi.org/10.1007/s41115-021-00013-z}
  {\bibfield  {journal} {\bibinfo  {journal} {Liv. Rev. Comput. Astrophys.}\
  }\textbf {\bibinfo {volume} {8}},\ \bibinfo {pages} {1} (\bibinfo {year}
  {2022})},\ \Eprint {https://arxiv.org/abs/2112.05165} {arXiv:2112.05165
  [astro-ph.CO]} \BibitemShut {NoStop}%
\bibitem [{\citenamefont {List}\ \emph {et~al.}(2024)\citenamefont {List},
  \citenamefont {Hahn},\ and\ \citenamefont {Rampf}}]{List:2023kbb}%
  \BibitemOpen
  \bibfield  {author} {\bibinfo {author} {\bibfnamefont {F.}~\bibnamefont
  {List}}, \bibinfo {author} {\bibfnamefont {O.}~\bibnamefont {Hahn}},\ and\
  \bibinfo {author} {\bibfnamefont {C.}~\bibnamefont {Rampf}},\ }\bibfield
  {title} {\bibinfo {title} {{Starting Cosmological Simulations from the Big
  Bang}},\ }\href {https://doi.org/10.1103/PhysRevLett.132.131003} {\bibfield
  {journal} {\bibinfo  {journal} {Phys. Rev. Lett.}\ }\textbf {\bibinfo
  {volume} {132}},\ \bibinfo {pages} {131003} (\bibinfo {year} {2024})},\
  \Eprint {https://arxiv.org/abs/2309.10865} {arXiv:2309.10865 [astro-ph.CO]}
  \BibitemShut {NoStop}%
\bibitem [{\citenamefont {Iliev}\ \emph {et~al.}(2006)\citenamefont {Iliev},
  \citenamefont {Mellema}, \citenamefont {Pen}, \citenamefont {Merz},
  \citenamefont {Shapiro},\ and\ \citenamefont {Alvarez}}]{Iliev:2005sz}%
  \BibitemOpen
  \bibfield  {author} {\bibinfo {author} {\bibfnamefont {I.~T.}\ \bibnamefont
  {Iliev}}, \bibinfo {author} {\bibfnamefont {G.}~\bibnamefont {Mellema}},
  \bibinfo {author} {\bibfnamefont {U.-L.}\ \bibnamefont {Pen}}, \bibinfo
  {author} {\bibfnamefont {H.}~\bibnamefont {Merz}}, \bibinfo {author}
  {\bibfnamefont {P.~R.}\ \bibnamefont {Shapiro}},\ and\ \bibinfo {author}
  {\bibfnamefont {M.~A.}\ \bibnamefont {Alvarez}},\ }\bibfield  {title}
  {\bibinfo {title} {{Simulating cosmic reionization at large scales. 1. the
  geometry of reionization}},\ }\href
  {https://doi.org/10.1111/j.1365-2966.2006.10502.x} {\bibfield  {journal}
  {\bibinfo  {journal} {Mon. Not. Roy. Astron. Soc.}\ }\textbf {\bibinfo
  {volume} {369}},\ \bibinfo {pages} {1625} (\bibinfo {year} {2006})},\ \Eprint
  {https://arxiv.org/abs/astro-ph/0512187} {arXiv:astro-ph/0512187}
  \BibitemShut {NoStop}%
\bibitem [{\citenamefont {Mesinger}\ \emph {et~al.}(2011)\citenamefont
  {Mesinger}, \citenamefont {Furlanetto},\ and\ \citenamefont
  {Cen}}]{Mesinger:2010ne}%
  \BibitemOpen
  \bibfield  {author} {\bibinfo {author} {\bibfnamefont {A.}~\bibnamefont
  {Mesinger}}, \bibinfo {author} {\bibfnamefont {S.}~\bibnamefont
  {Furlanetto}},\ and\ \bibinfo {author} {\bibfnamefont {R.}~\bibnamefont
  {Cen}},\ }\bibfield  {title} {\bibinfo {title} {{21cmFAST: A Fast,
  Semi-Numerical Simulation of the High-Redshift 21-cm Signal}},\ }\href
  {https://doi.org/10.1111/j.1365-2966.2010.17731.x} {\bibfield  {journal}
  {\bibinfo  {journal} {Mon. Not. Roy. Astron. Soc.}\ }\textbf {\bibinfo
  {volume} {411}},\ \bibinfo {pages} {955} (\bibinfo {year} {2011})},\ \Eprint
  {https://arxiv.org/abs/1003.3878} {arXiv:1003.3878 [astro-ph.CO]}
  \BibitemShut {NoStop}%
\bibitem [{\citenamefont {Vogelsberger}\ \emph {et~al.}(2014)\citenamefont
  {Vogelsberger}, \citenamefont {Genel}, \citenamefont {Springel},
  \citenamefont {Torrey}, \citenamefont {Sijacki}, \citenamefont {Xu},
  \citenamefont {Snyder}, \citenamefont {Nelson},\ and\ \citenamefont
  {Hernquist}}]{Vogelsberger:2014dza}%
  \BibitemOpen
  \bibfield  {author} {\bibinfo {author} {\bibfnamefont {M.}~\bibnamefont
  {Vogelsberger}}, \bibinfo {author} {\bibfnamefont {S.}~\bibnamefont {Genel}},
  \bibinfo {author} {\bibfnamefont {V.}~\bibnamefont {Springel}}, \bibinfo
  {author} {\bibfnamefont {P.}~\bibnamefont {Torrey}}, \bibinfo {author}
  {\bibfnamefont {D.}~\bibnamefont {Sijacki}}, \bibinfo {author} {\bibfnamefont
  {D.}~\bibnamefont {Xu}}, \bibinfo {author} {\bibfnamefont {G.~F.}\
  \bibnamefont {Snyder}}, \bibinfo {author} {\bibfnamefont {D.}~\bibnamefont
  {Nelson}},\ and\ \bibinfo {author} {\bibfnamefont {L.}~\bibnamefont
  {Hernquist}},\ }\bibfield  {title} {\bibinfo {title} {{Introducing the
  Illustris Project: Simulating the coevolution of dark and visible matter in
  the Universe}},\ }\href {https://doi.org/10.1093/mnras/stu1536} {\bibfield
  {journal} {\bibinfo  {journal} {Mon. Not. Roy. Astron. Soc.}\ }\textbf
  {\bibinfo {volume} {444}},\ \bibinfo {pages} {1518} (\bibinfo {year}
  {2014})},\ \Eprint {https://arxiv.org/abs/1405.2921} {arXiv:1405.2921
  [astro-ph.CO]} \BibitemShut {NoStop}%
\bibitem [{\citenamefont {Molaro}\ \emph {et~al.}(2019)\citenamefont {Molaro},
  \citenamefont {Dav{\'e}}, \citenamefont {Hassan}, \citenamefont {Santos},\
  and\ \citenamefont {Finlator}}]{Molaro:2019mew}%
  \BibitemOpen
  \bibfield  {author} {\bibinfo {author} {\bibfnamefont {M.}~\bibnamefont
  {Molaro}}, \bibinfo {author} {\bibfnamefont {R.}~\bibnamefont {Dav{\'e}}},
  \bibinfo {author} {\bibfnamefont {S.}~\bibnamefont {Hassan}}, \bibinfo
  {author} {\bibfnamefont {M.~G.}\ \bibnamefont {Santos}},\ and\ \bibinfo
  {author} {\bibfnamefont {K.}~\bibnamefont {Finlator}},\ }\bibfield  {title}
  {\bibinfo {title} {{ARTIST: Fast radiative transfer for large-scale
  simulations of the epoch of reionisation}},\ }\href
  {https://doi.org/10.1093/mnras/stz2171} {\bibfield  {journal} {\bibinfo
  {journal} {Mon. Not. Roy. Astron. Soc.}\ }\textbf {\bibinfo {volume} {489}},\
  \bibinfo {pages} {5594} (\bibinfo {year} {2019})},\ \Eprint
  {https://arxiv.org/abs/1901.03340} {arXiv:1901.03340 [astro-ph.CO]}
  \BibitemShut {NoStop}%
\bibitem [{\citenamefont {Maity}\ and\ \citenamefont
  {Choudhury}(2022)}]{Maity:2021mhr}%
  \BibitemOpen
  \bibfield  {author} {\bibinfo {author} {\bibfnamefont {B.}~\bibnamefont
  {Maity}}\ and\ \bibinfo {author} {\bibfnamefont {T.~R.}\ \bibnamefont
  {Choudhury}},\ }\bibfield  {title} {\bibinfo {title} {{Probing the thermal
  history during reionization using a seminumerical photon-conserving code
  script}},\ }\href {https://doi.org/10.1093/mnras/stac182} {\bibfield
  {journal} {\bibinfo  {journal} {Mon. Not. Roy. Astron. Soc.}\ }\textbf
  {\bibinfo {volume} {511}},\ \bibinfo {pages} {2239} (\bibinfo {year}
  {2022})},\ \Eprint {https://arxiv.org/abs/2110.14231} {arXiv:2110.14231
  [astro-ph.CO]} \BibitemShut {NoStop}%
\bibitem [{\citenamefont {Trac}\ \emph {et~al.}(2022)\citenamefont {Trac},
  \citenamefont {Chen}, \citenamefont {Holst}, \citenamefont {Alvarez},\ and\
  \citenamefont {Cen}}]{Trac:2021qbn}%
  \BibitemOpen
  \bibfield  {author} {\bibinfo {author} {\bibfnamefont {H.}~\bibnamefont
  {Trac}}, \bibinfo {author} {\bibfnamefont {N.}~\bibnamefont {Chen}}, \bibinfo
  {author} {\bibfnamefont {I.}~\bibnamefont {Holst}}, \bibinfo {author}
  {\bibfnamefont {M.~A.}\ \bibnamefont {Alvarez}},\ and\ \bibinfo {author}
  {\bibfnamefont {R.}~\bibnamefont {Cen}},\ }\bibfield  {title} {\bibinfo
  {title} {{AMBER: A Semi-numerical Abundance Matching Box for the Epoch of
  Reionization}},\ }\href {https://doi.org/10.3847/1538-4357/ac5116} {\bibfield
   {journal} {\bibinfo  {journal} {Astrophys. J.}\ }\textbf {\bibinfo {volume}
  {927}},\ \bibinfo {pages} {186} (\bibinfo {year} {2022})},\ \Eprint
  {https://arxiv.org/abs/2109.10375} {arXiv:2109.10375 [astro-ph.CO]}
  \BibitemShut {NoStop}%
\bibitem [{\citenamefont {Chen}\ \emph {et~al.}(2023)\citenamefont {Chen},
  \citenamefont {Trac}, \citenamefont {Mukherjee},\ and\ \citenamefont
  {Cen}}]{Chen:2022lhr}%
  \BibitemOpen
  \bibfield  {author} {\bibinfo {author} {\bibfnamefont {N.}~\bibnamefont
  {Chen}}, \bibinfo {author} {\bibfnamefont {H.}~\bibnamefont {Trac}}, \bibinfo
  {author} {\bibfnamefont {S.}~\bibnamefont {Mukherjee}},\ and\ \bibinfo
  {author} {\bibfnamefont {R.}~\bibnamefont {Cen}},\ }\bibfield  {title}
  {\bibinfo {title} {{Patchy Kinetic
  Sunyaev{\textendash}Zel{\textquoteright}dovich Effect with Controlled
  Reionization History and Morphology}},\ }\href
  {https://doi.org/10.3847/1538-4357/ac8481} {\bibfield  {journal} {\bibinfo
  {journal} {Astrophys. J.}\ }\textbf {\bibinfo {volume} {943}},\ \bibinfo
  {pages} {138} (\bibinfo {year} {2023})},\ \Eprint
  {https://arxiv.org/abs/2203.04337} {arXiv:2203.04337 [astro-ph.CO]}
  \BibitemShut {NoStop}%
\bibitem [{\citenamefont {Gnedin}\ and\ \citenamefont
  {Madau}(2022)}]{Gnedin:2022eza}%
  \BibitemOpen
  \bibfield  {author} {\bibinfo {author} {\bibfnamefont {N.~Y.}\ \bibnamefont
  {Gnedin}}\ and\ \bibinfo {author} {\bibfnamefont {P.}~\bibnamefont {Madau}},\
  }\bibfield  {title} {\bibinfo {title} {{Modeling Cosmic Reionization}},\
  }\href@noop {} {\  (\bibinfo {year} {2022})},\ \Eprint
  {https://arxiv.org/abs/2208.02260} {arXiv:2208.02260 [astro-ph.CO]}
  \BibitemShut {NoStop}%
\bibitem [{\citenamefont {Sugiyama}(1979)}]{Sugiyama:1979mi}%
  \BibitemOpen
  \bibfield  {author} {\bibinfo {author} {\bibfnamefont {T.}~\bibnamefont
  {Sugiyama}},\ }\bibfield  {title} {\bibinfo {title} {{KINK - ANTIKINK
  COLLISIONS IN THE TWO-DIMENSIONAL PHI**4 MODEL}},\ }\href
  {https://doi.org/10.1143/PTP.61.1550} {\bibfield  {journal} {\bibinfo
  {journal} {Prog. Theor. Phys.}\ }\textbf {\bibinfo {volume} {61}},\ \bibinfo
  {pages} {1550} (\bibinfo {year} {1979})}\BibitemShut {NoStop}%
\bibitem [{\citenamefont {Matzner}(1988)}]{Matzner:1988qqj}%
  \BibitemOpen
  \bibfield  {author} {\bibinfo {author} {\bibfnamefont {R.~A.}\ \bibnamefont
  {Matzner}},\ }\bibfield  {title} {\bibinfo {title} {{Interaction of U(1)
  cosmic strings: Numerical intercommutation}},\ }\href
  {https://doi.org/10.1063/1.168306} {\bibfield  {journal} {\bibinfo  {journal}
  {Comput. Phys.}\ }\textbf {\bibinfo {volume} {2}},\ \bibinfo {pages} {51}
  (\bibinfo {year} {1988})}\BibitemShut {NoStop}%
\bibitem [{\citenamefont {Lee}\ and\ \citenamefont {Pang}(1992)}]{Lee:1991ax}%
  \BibitemOpen
  \bibfield  {author} {\bibinfo {author} {\bibfnamefont {T.~D.}\ \bibnamefont
  {Lee}}\ and\ \bibinfo {author} {\bibfnamefont {Y.}~\bibnamefont {Pang}},\
  }\bibfield  {title} {\bibinfo {title} {{Nontopological solitons}},\ }\href
  {https://doi.org/10.1016/0370-1573(92)90064-7} {\bibfield  {journal}
  {\bibinfo  {journal} {Phys. Rept.}\ }\textbf {\bibinfo {volume} {221}},\
  \bibinfo {pages} {251} (\bibinfo {year} {1992})}\BibitemShut {NoStop}%
\bibitem [{\citenamefont {Manton}\ and\ \citenamefont
  {Sutcliffe}(2004)}]{Manton:2004tk}%
  \BibitemOpen
  \bibfield  {author} {\bibinfo {author} {\bibfnamefont {N.~S.}\ \bibnamefont
  {Manton}}\ and\ \bibinfo {author} {\bibfnamefont {P.}~\bibnamefont
  {Sutcliffe}},\ }\href {https://doi.org/10.1017/CBO9780511617034} {\emph
  {\bibinfo {title} {{Topological solitons}}}},\ Cambridge Monographs on
  Mathematical Physics\ (\bibinfo  {publisher} {Cambridge University Press},\
  \bibinfo {year} {2004})\BibitemShut {NoStop}%
\bibitem [{\citenamefont {Vachaspati}(2007)}]{Vachaspati:2006zz}%
  \BibitemOpen
  \bibfield  {author} {\bibinfo {author} {\bibfnamefont {T.}~\bibnamefont
  {Vachaspati}},\ }\href {https://doi.org/10.1017/9781009290456} {\emph
  {\bibinfo {title} {{Kinks and Domain Walls : An Introduction to Classical and
  Quantum Solitons}}}}\ (\bibinfo  {publisher} {Oxford University Press},\
  \bibinfo {year} {2007})\BibitemShut {NoStop}%
\bibitem [{\citenamefont {Achucarro}\ and\ \citenamefont
  {de~Putter}(2006)}]{Achucarro:2006es}%
  \BibitemOpen
  \bibfield  {author} {\bibinfo {author} {\bibfnamefont {A.}~\bibnamefont
  {Achucarro}}\ and\ \bibinfo {author} {\bibfnamefont {R.}~\bibnamefont
  {de~Putter}},\ }\bibfield  {title} {\bibinfo {title} {{Effective
  non-intercommutation of local cosmic strings at high collision speeds}},\
  }\href {https://doi.org/10.1103/PhysRevD.74.121701} {\bibfield  {journal}
  {\bibinfo  {journal} {Phys. Rev. D}\ }\textbf {\bibinfo {volume} {74}},\
  \bibinfo {pages} {121701} (\bibinfo {year} {2006})},\ \Eprint
  {https://arxiv.org/abs/hep-th/0605084} {arXiv:hep-th/0605084} \BibitemShut
  {NoStop}%
\bibitem [{\citenamefont {Dorey}\ \emph {et~al.}(2011)\citenamefont {Dorey},
  \citenamefont {Mersh}, \citenamefont {Romanczukiewicz},\ and\ \citenamefont
  {Shnir}}]{Dorey:2011yw}%
  \BibitemOpen
  \bibfield  {author} {\bibinfo {author} {\bibfnamefont {P.}~\bibnamefont
  {Dorey}}, \bibinfo {author} {\bibfnamefont {K.}~\bibnamefont {Mersh}},
  \bibinfo {author} {\bibfnamefont {T.}~\bibnamefont {Romanczukiewicz}},\ and\
  \bibinfo {author} {\bibfnamefont {Y.}~\bibnamefont {Shnir}},\ }\bibfield
  {title} {\bibinfo {title} {{Kink-antikink collisions in the $\phi^6$
  model}},\ }\href {https://doi.org/10.1103/PhysRevLett.107.091602} {\bibfield
  {journal} {\bibinfo  {journal} {Phys. Rev. Lett.}\ }\textbf {\bibinfo
  {volume} {107}},\ \bibinfo {pages} {091602} (\bibinfo {year} {2011})},\
  \Eprint {https://arxiv.org/abs/1101.5951} {arXiv:1101.5951 [hep-th]}
  \BibitemShut {NoStop}%
\bibitem [{\citenamefont {Gani}\ \emph {et~al.}(2014)\citenamefont {Gani},
  \citenamefont {Kudryavtsev},\ and\ \citenamefont {Lizunova}}]{Gani:2014gxa}%
  \BibitemOpen
  \bibfield  {author} {\bibinfo {author} {\bibfnamefont {V.~A.}\ \bibnamefont
  {Gani}}, \bibinfo {author} {\bibfnamefont {A.~E.}\ \bibnamefont
  {Kudryavtsev}},\ and\ \bibinfo {author} {\bibfnamefont {M.~A.}\ \bibnamefont
  {Lizunova}},\ }\bibfield  {title} {\bibinfo {title} {{Kink interactions in
  the (1+1)-dimensional $\phi^6$ model}},\ }\href
  {https://doi.org/10.1103/PhysRevD.89.125009} {\bibfield  {journal} {\bibinfo
  {journal} {Phys. Rev. D}\ }\textbf {\bibinfo {volume} {89}},\ \bibinfo
  {pages} {125009} (\bibinfo {year} {2014})},\ \Eprint
  {https://arxiv.org/abs/1402.5903} {arXiv:1402.5903 [hep-th]} \BibitemShut
  {NoStop}%
\bibitem [{\citenamefont {Amin}\ \emph {et~al.}(2014)\citenamefont {Amin},
  \citenamefont {Banik}, \citenamefont {Negreanu},\ and\ \citenamefont
  {Yang}}]{Amin:2014fua}%
  \BibitemOpen
  \bibfield  {author} {\bibinfo {author} {\bibfnamefont {M.~A.}\ \bibnamefont
  {Amin}}, \bibinfo {author} {\bibfnamefont {I.}~\bibnamefont {Banik}},
  \bibinfo {author} {\bibfnamefont {C.}~\bibnamefont {Negreanu}},\ and\
  \bibinfo {author} {\bibfnamefont {I.-S.}\ \bibnamefont {Yang}},\ }\bibfield
  {title} {\bibinfo {title} {{Ultrarelativistic oscillon collisions}},\ }\href
  {https://doi.org/10.1103/PhysRevD.90.085024} {\bibfield  {journal} {\bibinfo
  {journal} {Phys. Rev. D}\ }\textbf {\bibinfo {volume} {90}},\ \bibinfo
  {pages} {085024} (\bibinfo {year} {2014})},\ \Eprint
  {https://arxiv.org/abs/1410.1822} {arXiv:1410.1822 [hep-th]} \BibitemShut
  {NoStop}%
\bibitem [{\citenamefont {Amin}\ \emph
  {et~al.}(2013{\natexlab{a}})\citenamefont {Amin}, \citenamefont {Lim},\ and\
  \citenamefont {Yang}}]{Amin:2013dqa}%
  \BibitemOpen
  \bibfield  {author} {\bibinfo {author} {\bibfnamefont {M.~A.}\ \bibnamefont
  {Amin}}, \bibinfo {author} {\bibfnamefont {E.~A.}\ \bibnamefont {Lim}},\ and\
  \bibinfo {author} {\bibfnamefont {I.-S.}\ \bibnamefont {Yang}},\ }\bibfield
  {title} {\bibinfo {title} {{Clash of Kinks: Phase Shifts in Colliding
  Nonintegrable Solitons}},\ }\href
  {https://doi.org/10.1103/PhysRevLett.111.224101} {\bibfield  {journal}
  {\bibinfo  {journal} {Phys. Rev. Lett.}\ }\textbf {\bibinfo {volume} {111}},\
  \bibinfo {pages} {224101} (\bibinfo {year} {2013}{\natexlab{a}})},\ \Eprint
  {https://arxiv.org/abs/1308.0605} {arXiv:1308.0605 [hep-th]} \BibitemShut
  {NoStop}%
\bibitem [{\citenamefont {Amin}\ \emph
  {et~al.}(2013{\natexlab{b}})\citenamefont {Amin}, \citenamefont {Lim},\ and\
  \citenamefont {Yang}}]{Amin:2013eqa}%
  \BibitemOpen
  \bibfield  {author} {\bibinfo {author} {\bibfnamefont {M.~A.}\ \bibnamefont
  {Amin}}, \bibinfo {author} {\bibfnamefont {E.~A.}\ \bibnamefont {Lim}},\ and\
  \bibinfo {author} {\bibfnamefont {I.-S.}\ \bibnamefont {Yang}},\ }\bibfield
  {title} {\bibinfo {title} {{A scattering theory of ultrarelativistic
  solitons}},\ }\href {https://doi.org/10.1103/PhysRevD.88.105024} {\bibfield
  {journal} {\bibinfo  {journal} {Phys. Rev. D}\ }\textbf {\bibinfo {volume}
  {88}},\ \bibinfo {pages} {105024} (\bibinfo {year} {2013}{\natexlab{b}})},\
  \Eprint {https://arxiv.org/abs/1308.0606} {arXiv:1308.0606 [hep-th]}
  \BibitemShut {NoStop}%
\bibitem [{\citenamefont {Kinach}\ and\ \citenamefont
  {Choptuik}(2024{\natexlab{a}})}]{Kinach:2024hfa}%
  \BibitemOpen
  \bibfield  {author} {\bibinfo {author} {\bibfnamefont {M.~P.}\ \bibnamefont
  {Kinach}}\ and\ \bibinfo {author} {\bibfnamefont {M.~W.}\ \bibnamefont
  {Choptuik}},\ }\bibfield  {title} {\bibinfo {title} {{Relativistic head-on
  collisions of U(1) gauged Q-balls}},\ }\href
  {https://doi.org/10.1103/PhysRevD.110.015012} {\bibfield  {journal} {\bibinfo
   {journal} {Phys. Rev. D}\ }\textbf {\bibinfo {volume} {110}},\ \bibinfo
  {pages} {015012} (\bibinfo {year} {2024}{\natexlab{a}})},\ \Eprint
  {https://arxiv.org/abs/2404.04323} {arXiv:2404.04323 [hep-th]} \BibitemShut
  {NoStop}%
\bibitem [{\citenamefont {Kinach}\ and\ \citenamefont
  {Choptuik}(2024{\natexlab{b}})}]{Kinach:2024qzc}%
  \BibitemOpen
  \bibfield  {author} {\bibinfo {author} {\bibfnamefont {M.~P.}\ \bibnamefont
  {Kinach}}\ and\ \bibinfo {author} {\bibfnamefont {M.~W.}\ \bibnamefont
  {Choptuik}},\ }\bibfield  {title} {\bibinfo {title} {{Dynamics of U(1) gauged
  Q-balls in three spatial dimensions}},\ }\href
  {https://doi.org/10.1103/PhysRevD.110.075033} {\bibfield  {journal} {\bibinfo
   {journal} {Phys. Rev. D}\ }\textbf {\bibinfo {volume} {110}},\ \bibinfo
  {pages} {075033} (\bibinfo {year} {2024}{\natexlab{b}})},\ \Eprint
  {https://arxiv.org/abs/2408.07561} {arXiv:2408.07561 [hep-th]} \BibitemShut
  {NoStop}%
\bibitem [{\citenamefont {Vredevoogd}\ and\ \citenamefont
  {Pratt}(2009)}]{Vredevoogd:2009zu}%
  \BibitemOpen
  \bibfield  {author} {\bibinfo {author} {\bibfnamefont {J.}~\bibnamefont
  {Vredevoogd}}\ and\ \bibinfo {author} {\bibfnamefont {S.}~\bibnamefont
  {Pratt}},\ }\bibfield  {title} {\bibinfo {title} {{Universal Flow in the
  First fm/c at RHIC}},\ }\href
  {https://doi.org/10.1016/j.nuclphysa.2009.10.140} {\bibfield  {journal}
  {\bibinfo  {journal} {Nucl. Phys. A}\ }\textbf {\bibinfo {volume} {830}},\
  \bibinfo {pages} {515C} (\bibinfo {year} {2009})},\ \Eprint
  {https://arxiv.org/abs/0907.4733} {arXiv:0907.4733 [nucl-th]} \BibitemShut
  {NoStop}%
\bibitem [{\citenamefont {Schenke}\ \emph {et~al.}(2012)\citenamefont
  {Schenke}, \citenamefont {Tribedy},\ and\ \citenamefont
  {Venugopalan}}]{Schenke:2012wb}%
  \BibitemOpen
  \bibfield  {author} {\bibinfo {author} {\bibfnamefont {B.}~\bibnamefont
  {Schenke}}, \bibinfo {author} {\bibfnamefont {P.}~\bibnamefont {Tribedy}},\
  and\ \bibinfo {author} {\bibfnamefont {R.}~\bibnamefont {Venugopalan}},\
  }\bibfield  {title} {\bibinfo {title} {{Fluctuating Glasma initial conditions
  and flow in heavy ion collisions}},\ }\href
  {https://doi.org/10.1103/PhysRevLett.108.252301} {\bibfield  {journal}
  {\bibinfo  {journal} {Phys. Rev. Lett.}\ }\textbf {\bibinfo {volume} {108}},\
  \bibinfo {pages} {252301} (\bibinfo {year} {2012})},\ \Eprint
  {https://arxiv.org/abs/1202.6646} {arXiv:1202.6646 [nucl-th]} \BibitemShut
  {NoStop}%
\bibitem [{\citenamefont {Gale}\ \emph {et~al.}(2013)\citenamefont {Gale},
  \citenamefont {Jeon},\ and\ \citenamefont {Schenke}}]{Gale:2013da}%
  \BibitemOpen
  \bibfield  {author} {\bibinfo {author} {\bibfnamefont {C.}~\bibnamefont
  {Gale}}, \bibinfo {author} {\bibfnamefont {S.}~\bibnamefont {Jeon}},\ and\
  \bibinfo {author} {\bibfnamefont {B.}~\bibnamefont {Schenke}},\ }\bibfield
  {title} {\bibinfo {title} {{Hydrodynamic Modeling of Heavy-Ion Collisions}},\
  }\href {https://doi.org/10.1142/S0217751X13400113} {\bibfield  {journal}
  {\bibinfo  {journal} {Int. J. Mod. Phys. A}\ }\textbf {\bibinfo {volume}
  {28}},\ \bibinfo {pages} {1340011} (\bibinfo {year} {2013})},\ \Eprint
  {https://arxiv.org/abs/1301.5893} {arXiv:1301.5893 [nucl-th]} \BibitemShut
  {NoStop}%
\bibitem [{\citenamefont {Heinz}\ and\ \citenamefont
  {Snellings}(2013)}]{Heinz:2013th}%
  \BibitemOpen
  \bibfield  {author} {\bibinfo {author} {\bibfnamefont {U.}~\bibnamefont
  {Heinz}}\ and\ \bibinfo {author} {\bibfnamefont {R.}~\bibnamefont
  {Snellings}},\ }\bibfield  {title} {\bibinfo {title} {{Collective flow and
  viscosity in relativistic heavy-ion collisions}},\ }\href
  {https://doi.org/10.1146/annurev-nucl-102212-170540} {\bibfield  {journal}
  {\bibinfo  {journal} {Ann. Rev. Nucl. Part. Sci.}\ }\textbf {\bibinfo
  {volume} {63}},\ \bibinfo {pages} {123} (\bibinfo {year} {2013})},\ \Eprint
  {https://arxiv.org/abs/1301.2826} {arXiv:1301.2826 [nucl-th]} \BibitemShut
  {NoStop}%
\bibitem [{\citenamefont {Moreland}\ \emph {et~al.}(2015)\citenamefont
  {Moreland}, \citenamefont {Bernhard},\ and\ \citenamefont
  {Bass}}]{Moreland:2014oya}%
  \BibitemOpen
  \bibfield  {author} {\bibinfo {author} {\bibfnamefont {J.~S.}\ \bibnamefont
  {Moreland}}, \bibinfo {author} {\bibfnamefont {J.~E.}\ \bibnamefont
  {Bernhard}},\ and\ \bibinfo {author} {\bibfnamefont {S.~A.}\ \bibnamefont
  {Bass}},\ }\bibfield  {title} {\bibinfo {title} {{Alternative ansatz to
  wounded nucleon and binary collision scaling in high-energy nuclear
  collisions}},\ }\href {https://doi.org/10.1103/PhysRevC.92.011901} {\bibfield
   {journal} {\bibinfo  {journal} {Phys. Rev. C}\ }\textbf {\bibinfo {volume}
  {92}},\ \bibinfo {pages} {011901} (\bibinfo {year} {2015})},\ \Eprint
  {https://arxiv.org/abs/1412.4708} {arXiv:1412.4708 [nucl-th]} \BibitemShut
  {NoStop}%
\bibitem [{\citenamefont {Carrington}\ \emph {et~al.}(2022)\citenamefont
  {Carrington}, \citenamefont {Czajka},\ and\ \citenamefont
  {Mrowczynski}}]{Carrington:2020ssh}%
  \BibitemOpen
  \bibfield  {author} {\bibinfo {author} {\bibfnamefont {M.~E.}\ \bibnamefont
  {Carrington}}, \bibinfo {author} {\bibfnamefont {A.}~\bibnamefont {Czajka}},\
  and\ \bibinfo {author} {\bibfnamefont {S.}~\bibnamefont {Mrowczynski}},\
  }\bibfield  {title} {\bibinfo {title} {{The energy-momentum tensor at the
  earliest stage of relativistic heavy-ion collisions}},\ }\href
  {https://doi.org/10.1140/epja/s10050-021-00600-x} {\bibfield  {journal}
  {\bibinfo  {journal} {Eur. Phys. J. A}\ }\textbf {\bibinfo {volume} {58}},\
  \bibinfo {pages} {5} (\bibinfo {year} {2022})},\ \Eprint
  {https://arxiv.org/abs/2012.03042} {arXiv:2012.03042 [hep-ph]} \BibitemShut
  {NoStop}%
\bibitem [{\citenamefont {Joyce}\ \emph {et~al.}(2005)\citenamefont {Joyce},
  \citenamefont {Levesque},\ and\ \citenamefont {Marcos}}]{Joyce:2004em}%
  \BibitemOpen
  \bibfield  {author} {\bibinfo {author} {\bibfnamefont {M.}~\bibnamefont
  {Joyce}}, \bibinfo {author} {\bibfnamefont {D.}~\bibnamefont {Levesque}},\
  and\ \bibinfo {author} {\bibfnamefont {B.}~\bibnamefont {Marcos}},\
  }\bibfield  {title} {\bibinfo {title} {{A Method of generating initial
  conditions for cosmological N body simulations}},\ }\href
  {https://doi.org/10.1103/PhysRevD.72.103509} {\bibfield  {journal} {\bibinfo
  {journal} {Phys. Rev. D}\ }\textbf {\bibinfo {volume} {72}},\ \bibinfo
  {pages} {103509} (\bibinfo {year} {2005})},\ \Eprint
  {https://arxiv.org/abs/astro-ph/0411607} {arXiv:astro-ph/0411607}
  \BibitemShut {NoStop}%
\bibitem [{\citenamefont {Amin}\ \emph
  {et~al.}(2025{\natexlab{a}})\citenamefont {Amin}, \citenamefont {Delos},\
  and\ \citenamefont {Mirbabayi}}]{Amin:2025dtd}%
  \BibitemOpen
  \bibfield  {author} {\bibinfo {author} {\bibfnamefont {M.~A.}\ \bibnamefont
  {Amin}}, \bibinfo {author} {\bibfnamefont {M.~S.}\ \bibnamefont {Delos}},\
  and\ \bibinfo {author} {\bibfnamefont {M.}~\bibnamefont {Mirbabayi}},\
  }\bibfield  {title} {\bibinfo {title} {{Structure Formation with Warm White
  Noise: Effects of Finite Number Density and Velocity Dispersion in Particle
  and Wave Dark Matter}},\ }\href@noop {} {\  (\bibinfo {year}
  {2025}{\natexlab{a}})},\ \Eprint {https://arxiv.org/abs/2503.20881}
  {arXiv:2503.20881 [astro-ph.CO]} \BibitemShut {NoStop}%
\bibitem [{\citenamefont {Madelung}(1927)}]{Madelung:1927ksh}%
  \BibitemOpen
  \bibfield  {author} {\bibinfo {author} {\bibfnamefont {E.}~\bibnamefont
  {Madelung}},\ }\bibfield  {title} {\bibinfo {title} {{Quantentheorie in
  hydrodynamischer Form}},\ }\href {https://doi.org/10.1007/BF01400372}
  {\bibfield  {journal} {\bibinfo  {journal} {Z. Phys.}\ }\textbf {\bibinfo
  {volume} {40}},\ \bibinfo {pages} {322} (\bibinfo {year} {1927})}\BibitemShut
  {NoStop}%
\bibitem [{\citenamefont {Shellard}(1987)}]{Shellard:1987bv}%
  \BibitemOpen
  \bibfield  {author} {\bibinfo {author} {\bibfnamefont {E.~P.~S.}\
  \bibnamefont {Shellard}},\ }\bibfield  {title} {\bibinfo {title} {{Cosmic
  String Interactions}},\ }\href {https://doi.org/10.1016/0550-3213(87)90290-2}
  {\bibfield  {journal} {\bibinfo  {journal} {Nucl. Phys. B}\ }\textbf
  {\bibinfo {volume} {283}},\ \bibinfo {pages} {624} (\bibinfo {year}
  {1987})}\BibitemShut {NoStop}%
\bibitem [{\citenamefont {Veltmaat}\ \emph {et~al.}(2018)\citenamefont
  {Veltmaat}, \citenamefont {Niemeyer},\ and\ \citenamefont
  {Schwabe}}]{Veltmaat:2018dfz}%
  \BibitemOpen
  \bibfield  {author} {\bibinfo {author} {\bibfnamefont {J.}~\bibnamefont
  {Veltmaat}}, \bibinfo {author} {\bibfnamefont {J.~C.}\ \bibnamefont
  {Niemeyer}},\ and\ \bibinfo {author} {\bibfnamefont {B.}~\bibnamefont
  {Schwabe}},\ }\bibfield  {title} {\bibinfo {title} {{Formation and structure
  of ultralight bosonic dark matter halos}},\ }\href
  {https://doi.org/10.1103/PhysRevD.98.043509} {\bibfield  {journal} {\bibinfo
  {journal} {Phys. Rev. D}\ }\textbf {\bibinfo {volume} {98}},\ \bibinfo
  {pages} {043509} (\bibinfo {year} {2018})},\ \Eprint
  {https://arxiv.org/abs/1804.09647} {arXiv:1804.09647 [astro-ph.CO]}
  \BibitemShut {NoStop}%
\bibitem [{\citenamefont {Figueroa}\ \emph {et~al.}(2021)\citenamefont
  {Figueroa}, \citenamefont {Florio}, \citenamefont {Torrenti},\ and\
  \citenamefont {Valkenburg}}]{Figueroa:2020rrl}%
  \BibitemOpen
  \bibfield  {author} {\bibinfo {author} {\bibfnamefont {D.~G.}\ \bibnamefont
  {Figueroa}}, \bibinfo {author} {\bibfnamefont {A.}~\bibnamefont {Florio}},
  \bibinfo {author} {\bibfnamefont {F.}~\bibnamefont {Torrenti}},\ and\
  \bibinfo {author} {\bibfnamefont {W.}~\bibnamefont {Valkenburg}},\ }\bibfield
   {title} {\bibinfo {title} {{The art of simulating the early Universe -- Part
  I}},\ }\href {https://doi.org/10.1088/1475-7516/2021/04/035} {\bibfield
  {journal} {\bibinfo  {journal} {JCAP}\ }\textbf {\bibinfo {volume} {04}},\
  \bibinfo {pages} {035}},\ \Eprint {https://arxiv.org/abs/2006.15122}
  {arXiv:2006.15122 [astro-ph.CO]} \BibitemShut {NoStop}%
\bibitem [{\citenamefont {May}\ and\ \citenamefont
  {Springel}(2023)}]{May:2022gus}%
  \BibitemOpen
  \bibfield  {author} {\bibinfo {author} {\bibfnamefont {S.}~\bibnamefont
  {May}}\ and\ \bibinfo {author} {\bibfnamefont {V.}~\bibnamefont {Springel}},\
  }\bibfield  {title} {\bibinfo {title} {{The halo mass function and filaments
  in full cosmological simulations with fuzzy dark matter}},\ }\href
  {https://doi.org/10.1093/mnras/stad2031} {\bibfield  {journal} {\bibinfo
  {journal} {Mon. Not. Roy. Astron. Soc.}\ }\textbf {\bibinfo {volume} {524}},\
  \bibinfo {pages} {4256} (\bibinfo {year} {2023})},\ \Eprint
  {https://arxiv.org/abs/2209.14886} {arXiv:2209.14886 [astro-ph.CO]}
  \BibitemShut {NoStop}%
\bibitem [{\citenamefont {Ling}\ and\ \citenamefont
  {Amin}(2025)}]{Ling:2024qfv}%
  \BibitemOpen
  \bibfield  {author} {\bibinfo {author} {\bibfnamefont {S.}~\bibnamefont
  {Ling}}\ and\ \bibinfo {author} {\bibfnamefont {M.~A.}\ \bibnamefont
  {Amin}},\ }\bibfield  {title} {\bibinfo {title} {{Free streaming in warm wave
  dark matter}},\ }\href {https://doi.org/10.1088/1475-7516/2025/02/025}
  {\bibfield  {journal} {\bibinfo  {journal} {JCAP}\ }\textbf {\bibinfo
  {volume} {02}},\ \bibinfo {pages} {025}},\ \Eprint
  {https://arxiv.org/abs/2408.05591} {arXiv:2408.05591 [astro-ph.CO]}
  \BibitemShut {NoStop}%
\bibitem [{\citenamefont {Barone}\ \emph {et~al.}(1971)\citenamefont {Barone},
  \citenamefont {Esposito}, \citenamefont {Magee},\ and\ \citenamefont
  {Scott}}]{Barone1971}%
  \BibitemOpen
  \bibfield  {author} {\bibinfo {author} {\bibfnamefont {A.}~\bibnamefont
  {Barone}}, \bibinfo {author} {\bibfnamefont {F.}~\bibnamefont {Esposito}},
  \bibinfo {author} {\bibfnamefont {C.~J.}\ \bibnamefont {Magee}},\ and\
  \bibinfo {author} {\bibfnamefont {A.~C.}\ \bibnamefont {Scott}},\ }\bibfield
  {title} {\bibinfo {title} {Theory and applications of the sine-gordon
  equation},\ }\href {https://doi.org/10.1007/bf02820622} {\bibfield  {journal}
  {\bibinfo  {journal} {La Rivista del Nuovo Cimento}\ }\textbf {\bibinfo
  {volume} {1}},\ \bibinfo {pages} {227–267} (\bibinfo {year}
  {1971})}\BibitemShut {NoStop}%
\bibitem [{\citenamefont {Coleman}(1975)}]{Coleman:1974bu}%
  \BibitemOpen
  \bibfield  {author} {\bibinfo {author} {\bibfnamefont {S.~R.}\ \bibnamefont
  {Coleman}},\ }\bibfield  {title} {\bibinfo {title} {{The Quantum Sine-Gordon
  Equation as the Massive Thirring Model}},\ }\href
  {https://doi.org/10.1103/PhysRevD.11.2088} {\bibfield  {journal} {\bibinfo
  {journal} {Phys. Rev. D}\ }\textbf {\bibinfo {volume} {11}},\ \bibinfo
  {pages} {2088} (\bibinfo {year} {1975})}\BibitemShut {NoStop}%
\bibitem [{\citenamefont {Mandelstam}(1975)}]{Mandelstam:1975hb}%
  \BibitemOpen
  \bibfield  {author} {\bibinfo {author} {\bibfnamefont {S.}~\bibnamefont
  {Mandelstam}},\ }\bibfield  {title} {\bibinfo {title} {{Soliton Operators for
  the Quantized Sine-Gordon Equation}},\ }\href
  {https://doi.org/10.1103/PhysRevD.11.3026} {\bibfield  {journal} {\bibinfo
  {journal} {Phys. Rev. D}\ }\textbf {\bibinfo {volume} {11}},\ \bibinfo
  {pages} {3026} (\bibinfo {year} {1975})}\BibitemShut {NoStop}%
\bibitem [{\citenamefont {Callan}(1982)}]{Callan:1982au}%
  \BibitemOpen
  \bibfield  {author} {\bibinfo {author} {\bibfnamefont {C.~G.}\ \bibnamefont
  {Callan}, \bibfnamefont {Jr.}},\ }\bibfield  {title} {\bibinfo {title}
  {{Dyon-Fermion Dynamics}},\ }\href {https://doi.org/10.1103/PhysRevD.26.2058}
  {\bibfield  {journal} {\bibinfo  {journal} {Phys. Rev. D}\ }\textbf {\bibinfo
  {volume} {26}},\ \bibinfo {pages} {2058} (\bibinfo {year}
  {1982})}\BibitemShut {NoStop}%
\bibitem [{\citenamefont {Malomed}(2014)}]{Malomed:2014eba}%
  \BibitemOpen
  \bibfield  {author} {\bibinfo {author} {\bibfnamefont {B.~A.}\ \bibnamefont
  {Malomed}},\ }\bibfield  {title} {\bibinfo {title} {{The sine-Gordon Model:
  General Background, Physical Motivations, Inverse Scattering, and
  Solitons}},\ }\href {https://doi.org/10.1007/978-3-319-06722-3_1} {\bibfield
  {journal} {\bibinfo  {journal} {Nonlinear Systems and Complexity}\ }\textbf
  {\bibinfo {volume} {10}},\ \bibinfo {pages} {1} (\bibinfo {year}
  {2014})}\BibitemShut {NoStop}%
\bibitem [{\citenamefont {Helfer}\ \emph {et~al.}(2019)\citenamefont {Helfer},
  \citenamefont {Lim}, \citenamefont {Garcia},\ and\ \citenamefont
  {Amin}}]{Helfer:2018vtq}%
  \BibitemOpen
  \bibfield  {author} {\bibinfo {author} {\bibfnamefont {T.}~\bibnamefont
  {Helfer}}, \bibinfo {author} {\bibfnamefont {E.~A.}\ \bibnamefont {Lim}},
  \bibinfo {author} {\bibfnamefont {M.~A.~G.}\ \bibnamefont {Garcia}},\ and\
  \bibinfo {author} {\bibfnamefont {M.~A.}\ \bibnamefont {Amin}},\ }\bibfield
  {title} {\bibinfo {title} {{Gravitational Wave Emission from Collisions of
  Compact Scalar Solitons}},\ }\href
  {https://doi.org/10.1103/PhysRevD.99.044046} {\bibfield  {journal} {\bibinfo
  {journal} {Phys. Rev. D}\ }\textbf {\bibinfo {volume} {99}},\ \bibinfo
  {pages} {044046} (\bibinfo {year} {2019})},\ \Eprint
  {https://arxiv.org/abs/1802.06733} {arXiv:1802.06733 [gr-qc]} \BibitemShut
  {NoStop}%
\bibitem [{\citenamefont {Eby}\ \emph {et~al.}(2017)\citenamefont {Eby},
  \citenamefont {Leembruggen}, \citenamefont {Leeney}, \citenamefont
  {Suranyi},\ and\ \citenamefont {Wijewardhana}}]{Eby:2017xaw}%
  \BibitemOpen
  \bibfield  {author} {\bibinfo {author} {\bibfnamefont {J.}~\bibnamefont
  {Eby}}, \bibinfo {author} {\bibfnamefont {M.}~\bibnamefont {Leembruggen}},
  \bibinfo {author} {\bibfnamefont {J.}~\bibnamefont {Leeney}}, \bibinfo
  {author} {\bibfnamefont {P.}~\bibnamefont {Suranyi}},\ and\ \bibinfo {author}
  {\bibfnamefont {L.~C.~R.}\ \bibnamefont {Wijewardhana}},\ }\bibfield  {title}
  {\bibinfo {title} {{Collisions of Dark Matter Axion Stars with Astrophysical
  Sources}},\ }\href {https://doi.org/10.1007/JHEP04(2017)099} {\bibfield
  {journal} {\bibinfo  {journal} {JHEP}\ }\textbf {\bibinfo {volume} {04}},\
  \bibinfo {pages} {099}},\ \Eprint {https://arxiv.org/abs/1701.01476}
  {arXiv:1701.01476 [astro-ph.CO]} \BibitemShut {NoStop}%
\bibitem [{\citenamefont {Widdicombe}\ \emph {et~al.}(2020)\citenamefont
  {Widdicombe}, \citenamefont {Helfer},\ and\ \citenamefont
  {Lim}}]{Widdicombe:2019woy}%
  \BibitemOpen
  \bibfield  {author} {\bibinfo {author} {\bibfnamefont {J.~Y.}\ \bibnamefont
  {Widdicombe}}, \bibinfo {author} {\bibfnamefont {T.}~\bibnamefont {Helfer}},\
  and\ \bibinfo {author} {\bibfnamefont {E.~A.}\ \bibnamefont {Lim}},\
  }\bibfield  {title} {\bibinfo {title} {{Black hole formation in relativistic
  Oscillaton collisions}},\ }\href
  {https://doi.org/10.1088/1475-7516/2020/01/027} {\bibfield  {journal}
  {\bibinfo  {journal} {JCAP}\ }\textbf {\bibinfo {volume} {01}},\ \bibinfo
  {pages} {027}},\ \Eprint {https://arxiv.org/abs/1910.01950} {arXiv:1910.01950
  [astro-ph.CO]} \BibitemShut {NoStop}%
\bibitem [{\citenamefont {Amin}\ and\ \citenamefont
  {Mocz}(2019)}]{Amin:2019ums}%
  \BibitemOpen
  \bibfield  {author} {\bibinfo {author} {\bibfnamefont {M.~A.}\ \bibnamefont
  {Amin}}\ and\ \bibinfo {author} {\bibfnamefont {P.}~\bibnamefont {Mocz}},\
  }\bibfield  {title} {\bibinfo {title} {{Formation, gravitational clustering,
  and interactions of nonrelativistic solitons in an expanding universe}},\
  }\href {https://doi.org/10.1103/PhysRevD.100.063507} {\bibfield  {journal}
  {\bibinfo  {journal} {Phys. Rev. D}\ }\textbf {\bibinfo {volume} {100}},\
  \bibinfo {pages} {063507} (\bibinfo {year} {2019})},\ \Eprint
  {https://arxiv.org/abs/1902.07261} {arXiv:1902.07261 [astro-ph.CO]}
  \BibitemShut {NoStop}%
\bibitem [{\citenamefont {Graham}\ \emph {et~al.}(2016)\citenamefont {Graham},
  \citenamefont {Mardon},\ and\ \citenamefont {Rajendran}}]{Graham:2015rva}%
  \BibitemOpen
  \bibfield  {author} {\bibinfo {author} {\bibfnamefont {P.~W.}\ \bibnamefont
  {Graham}}, \bibinfo {author} {\bibfnamefont {J.}~\bibnamefont {Mardon}},\
  and\ \bibinfo {author} {\bibfnamefont {S.}~\bibnamefont {Rajendran}},\
  }\bibfield  {title} {\bibinfo {title} {{Vector Dark Matter from Inflationary
  Fluctuations}},\ }\href {https://doi.org/10.1103/PhysRevD.93.103520}
  {\bibfield  {journal} {\bibinfo  {journal} {Phys. Rev. D}\ }\textbf {\bibinfo
  {volume} {93}},\ \bibinfo {pages} {103520} (\bibinfo {year} {2016})},\
  \Eprint {https://arxiv.org/abs/1504.02102} {arXiv:1504.02102 [hep-ph]}
  \BibitemShut {NoStop}%
\bibitem [{\citenamefont {Ahmed}\ \emph {et~al.}(2019)\citenamefont {Ahmed},
  \citenamefont {Grzadkowski},\ and\ \citenamefont {Socha}}]{Ahmed:2019mjo}%
  \BibitemOpen
  \bibfield  {author} {\bibinfo {author} {\bibfnamefont {A.}~\bibnamefont
  {Ahmed}}, \bibinfo {author} {\bibfnamefont {B.}~\bibnamefont {Grzadkowski}},\
  and\ \bibinfo {author} {\bibfnamefont {A.}~\bibnamefont {Socha}},\ }\bibfield
   {title} {\bibinfo {title} {{Production of Purely Gravitational Vector Dark
  Matter}},\ }\href {https://doi.org/10.5506/APhysPolB.50.1809} {\bibfield
  {journal} {\bibinfo  {journal} {Acta Phys. Polon. B}\ }\textbf {\bibinfo
  {volume} {50}},\ \bibinfo {pages} {1809} (\bibinfo {year}
  {2019})}\BibitemShut {NoStop}%
\bibitem [{\citenamefont {Ema}\ \emph {et~al.}(2019)\citenamefont {Ema},
  \citenamefont {Nakayama},\ and\ \citenamefont {Tang}}]{Ema:2019yrd}%
  \BibitemOpen
  \bibfield  {author} {\bibinfo {author} {\bibfnamefont {Y.}~\bibnamefont
  {Ema}}, \bibinfo {author} {\bibfnamefont {K.}~\bibnamefont {Nakayama}},\ and\
  \bibinfo {author} {\bibfnamefont {Y.}~\bibnamefont {Tang}},\ }\bibfield
  {title} {\bibinfo {title} {{Production of purely gravitational dark matter:
  the case of fermion and vector boson}},\ }\href
  {https://doi.org/10.1007/JHEP07(2019)060} {\bibfield  {journal} {\bibinfo
  {journal} {JHEP}\ }\textbf {\bibinfo {volume} {07}},\ \bibinfo {pages}
  {060}},\ \Eprint {https://arxiv.org/abs/1903.10973} {arXiv:1903.10973
  [hep-ph]} \BibitemShut {NoStop}%
\bibitem [{\citenamefont {Ahmed}\ \emph {et~al.}(2020)\citenamefont {Ahmed},
  \citenamefont {Grzadkowski},\ and\ \citenamefont {Socha}}]{Ahmed:2020fhc}%
  \BibitemOpen
  \bibfield  {author} {\bibinfo {author} {\bibfnamefont {A.}~\bibnamefont
  {Ahmed}}, \bibinfo {author} {\bibfnamefont {B.}~\bibnamefont {Grzadkowski}},\
  and\ \bibinfo {author} {\bibfnamefont {A.}~\bibnamefont {Socha}},\ }\bibfield
   {title} {\bibinfo {title} {{Gravitational production of vector dark
  matter}},\ }\href {https://doi.org/10.1007/JHEP08(2020)059} {\bibfield
  {journal} {\bibinfo  {journal} {JHEP}\ }\textbf {\bibinfo {volume} {08}},\
  \bibinfo {pages} {059}},\ \Eprint {https://arxiv.org/abs/2005.01766}
  {arXiv:2005.01766 [hep-ph]} \BibitemShut {NoStop}%
\bibitem [{\citenamefont {Kolb}\ and\ \citenamefont
  {Long}(2021)}]{Kolb:2020fwh}%
  \BibitemOpen
  \bibfield  {author} {\bibinfo {author} {\bibfnamefont {E.~W.}\ \bibnamefont
  {Kolb}}\ and\ \bibinfo {author} {\bibfnamefont {A.~J.}\ \bibnamefont
  {Long}},\ }\bibfield  {title} {\bibinfo {title} {{Completely dark photons
  from gravitational particle production during the inflationary era}},\ }\href
  {https://doi.org/10.1007/JHEP03(2021)283} {\bibfield  {journal} {\bibinfo
  {journal} {JHEP}\ }\textbf {\bibinfo {volume} {03}},\ \bibinfo {pages}
  {283}},\ \Eprint {https://arxiv.org/abs/2009.03828} {arXiv:2009.03828
  [astro-ph.CO]} \BibitemShut {NoStop}%
\bibitem [{\citenamefont {Duch}\ \emph {et~al.}(2018)\citenamefont {Duch},
  \citenamefont {Grzadkowski},\ and\ \citenamefont {Huang}}]{Duch:2017khv}%
  \BibitemOpen
  \bibfield  {author} {\bibinfo {author} {\bibfnamefont {M.}~\bibnamefont
  {Duch}}, \bibinfo {author} {\bibfnamefont {B.}~\bibnamefont {Grzadkowski}},\
  and\ \bibinfo {author} {\bibfnamefont {D.}~\bibnamefont {Huang}},\ }\bibfield
   {title} {\bibinfo {title} {{Strongly self-interacting vector dark matter via
  freeze-in}},\ }\href {https://doi.org/10.1007/JHEP01(2018)020} {\bibfield
  {journal} {\bibinfo  {journal} {JHEP}\ }\textbf {\bibinfo {volume} {01}},\
  \bibinfo {pages} {020}},\ \Eprint {https://arxiv.org/abs/1710.00320}
  {arXiv:1710.00320 [hep-ph]} \BibitemShut {NoStop}%
\bibitem [{\citenamefont {Barman}\ \emph {et~al.}(2020)\citenamefont {Barman},
  \citenamefont {Bhattacharya},\ and\ \citenamefont
  {Grzadkowski}}]{Barman:2020ifq}%
  \BibitemOpen
  \bibfield  {author} {\bibinfo {author} {\bibfnamefont {B.}~\bibnamefont
  {Barman}}, \bibinfo {author} {\bibfnamefont {S.}~\bibnamefont
  {Bhattacharya}},\ and\ \bibinfo {author} {\bibfnamefont {B.}~\bibnamefont
  {Grzadkowski}},\ }\bibfield  {title} {\bibinfo {title} {{Feebly coupled
  vector boson dark matter in effective theory}},\ }\href
  {https://doi.org/10.1007/JHEP12(2020)162} {\bibfield  {journal} {\bibinfo
  {journal} {JHEP}\ }\textbf {\bibinfo {volume} {12}},\ \bibinfo {pages}
  {162}},\ \Eprint {https://arxiv.org/abs/2009.07438} {arXiv:2009.07438
  [hep-ph]} \BibitemShut {NoStop}%
\bibitem [{\citenamefont {Barman}\ \emph {et~al.}(2022)\citenamefont {Barman},
  \citenamefont {Bernal}, \citenamefont {Das},\ and\ \citenamefont
  {Roshan}}]{Barman:2021qds}%
  \BibitemOpen
  \bibfield  {author} {\bibinfo {author} {\bibfnamefont {B.}~\bibnamefont
  {Barman}}, \bibinfo {author} {\bibfnamefont {N.}~\bibnamefont {Bernal}},
  \bibinfo {author} {\bibfnamefont {A.}~\bibnamefont {Das}},\ and\ \bibinfo
  {author} {\bibfnamefont {R.}~\bibnamefont {Roshan}},\ }\bibfield  {title}
  {\bibinfo {title} {{Non-minimally coupled vector boson dark matter}},\ }\href
  {https://doi.org/10.1088/1475-7516/2022/01/047} {\bibfield  {journal}
  {\bibinfo  {journal} {JCAP}\ }\textbf {\bibinfo {volume} {01}}\bibfield
  {number} {\bibinfo  {number} { (01)},\ \bibinfo {pages} {047}},\ }\Eprint
  {https://arxiv.org/abs/2108.13447} {arXiv:2108.13447 [hep-ph]} \BibitemShut
  {NoStop}%
\bibitem [{\citenamefont {Agrawal}\ \emph {et~al.}(2018)\citenamefont
  {Agrawal}, \citenamefont {Marques-Tavares},\ and\ \citenamefont
  {Xue}}]{Agrawal:2017eqm}%
  \BibitemOpen
  \bibfield  {author} {\bibinfo {author} {\bibfnamefont {P.}~\bibnamefont
  {Agrawal}}, \bibinfo {author} {\bibfnamefont {G.}~\bibnamefont
  {Marques-Tavares}},\ and\ \bibinfo {author} {\bibfnamefont {W.}~\bibnamefont
  {Xue}},\ }\bibfield  {title} {\bibinfo {title} {{Opening up the QCD axion
  window}},\ }\href {https://doi.org/10.1007/JHEP03(2018)049} {\bibfield
  {journal} {\bibinfo  {journal} {JHEP}\ }\textbf {\bibinfo {volume} {03}},\
  \bibinfo {pages} {049}},\ \Eprint {https://arxiv.org/abs/1708.05008}
  {arXiv:1708.05008 [hep-ph]} \BibitemShut {NoStop}%
\bibitem [{\citenamefont {Agrawal}\ \emph {et~al.}(2020)\citenamefont
  {Agrawal}, \citenamefont {Kitajima}, \citenamefont {Reece}, \citenamefont
  {Sekiguchi},\ and\ \citenamefont {Takahashi}}]{Agrawal:2018vin}%
  \BibitemOpen
  \bibfield  {author} {\bibinfo {author} {\bibfnamefont {P.}~\bibnamefont
  {Agrawal}}, \bibinfo {author} {\bibfnamefont {N.}~\bibnamefont {Kitajima}},
  \bibinfo {author} {\bibfnamefont {M.}~\bibnamefont {Reece}}, \bibinfo
  {author} {\bibfnamefont {T.}~\bibnamefont {Sekiguchi}},\ and\ \bibinfo
  {author} {\bibfnamefont {F.}~\bibnamefont {Takahashi}},\ }\bibfield  {title}
  {\bibinfo {title} {{Relic Abundance of Dark Photon Dark Matter}},\ }\href
  {https://doi.org/10.1016/j.physletb.2019.135136} {\bibfield  {journal}
  {\bibinfo  {journal} {Phys. Lett. B}\ }\textbf {\bibinfo {volume} {801}},\
  \bibinfo {pages} {135136} (\bibinfo {year} {2020})},\ \Eprint
  {https://arxiv.org/abs/1810.07188} {arXiv:1810.07188 [hep-ph]} \BibitemShut
  {NoStop}%
\bibitem [{\citenamefont {Dror}\ \emph {et~al.}(2019)\citenamefont {Dror},
  \citenamefont {Harigaya},\ and\ \citenamefont {Narayan}}]{Dror:2018pdh}%
  \BibitemOpen
  \bibfield  {author} {\bibinfo {author} {\bibfnamefont {J.~A.}\ \bibnamefont
  {Dror}}, \bibinfo {author} {\bibfnamefont {K.}~\bibnamefont {Harigaya}},\
  and\ \bibinfo {author} {\bibfnamefont {V.}~\bibnamefont {Narayan}},\
  }\bibfield  {title} {\bibinfo {title} {{Parametric Resonance Production of
  Ultralight Vector Dark Matter}},\ }\href
  {https://doi.org/10.1103/PhysRevD.99.035036} {\bibfield  {journal} {\bibinfo
  {journal} {Phys. Rev. D}\ }\textbf {\bibinfo {volume} {99}},\ \bibinfo
  {pages} {035036} (\bibinfo {year} {2019})},\ \Eprint
  {https://arxiv.org/abs/1810.07195} {arXiv:1810.07195 [hep-ph]} \BibitemShut
  {NoStop}%
\bibitem [{\citenamefont {Co}\ \emph {et~al.}(2019)\citenamefont {Co},
  \citenamefont {Pierce}, \citenamefont {Zhang},\ and\ \citenamefont
  {Zhao}}]{Co:2018lka}%
  \BibitemOpen
  \bibfield  {author} {\bibinfo {author} {\bibfnamefont {R.~T.}\ \bibnamefont
  {Co}}, \bibinfo {author} {\bibfnamefont {A.}~\bibnamefont {Pierce}}, \bibinfo
  {author} {\bibfnamefont {Z.}~\bibnamefont {Zhang}},\ and\ \bibinfo {author}
  {\bibfnamefont {Y.}~\bibnamefont {Zhao}},\ }\bibfield  {title} {\bibinfo
  {title} {{Dark Photon Dark Matter Produced by Axion Oscillations}},\ }\href
  {https://doi.org/10.1103/PhysRevD.99.075002} {\bibfield  {journal} {\bibinfo
  {journal} {Phys. Rev. D}\ }\textbf {\bibinfo {volume} {99}},\ \bibinfo
  {pages} {075002} (\bibinfo {year} {2019})},\ \Eprint
  {https://arxiv.org/abs/1810.07196} {arXiv:1810.07196 [hep-ph]} \BibitemShut
  {NoStop}%
\bibitem [{\citenamefont {Bastero-Gil}\ \emph {et~al.}(2019)\citenamefont
  {Bastero-Gil}, \citenamefont {Santiago}, \citenamefont {Ubaldi},\ and\
  \citenamefont {Vega-Morales}}]{Bastero-Gil:2018uel}%
  \BibitemOpen
  \bibfield  {author} {\bibinfo {author} {\bibfnamefont {M.}~\bibnamefont
  {Bastero-Gil}}, \bibinfo {author} {\bibfnamefont {J.}~\bibnamefont
  {Santiago}}, \bibinfo {author} {\bibfnamefont {L.}~\bibnamefont {Ubaldi}},\
  and\ \bibinfo {author} {\bibfnamefont {R.}~\bibnamefont {Vega-Morales}},\
  }\bibfield  {title} {\bibinfo {title} {{Vector dark matter production at the
  end of inflation}},\ }\href {https://doi.org/10.1088/1475-7516/2019/04/015}
  {\bibfield  {journal} {\bibinfo  {journal} {JCAP}\ }\textbf {\bibinfo
  {volume} {04}},\ \bibinfo {pages} {015}},\ \Eprint
  {https://arxiv.org/abs/1810.07208} {arXiv:1810.07208 [hep-ph]} \BibitemShut
  {NoStop}%
\bibitem [{\citenamefont {Salehian}\ \emph
  {et~al.}(2021{\natexlab{a}})\citenamefont {Salehian}, \citenamefont {Gorji},
  \citenamefont {Firouzjahi},\ and\ \citenamefont
  {Mukohyama}}]{Salehian:2020asa}%
  \BibitemOpen
  \bibfield  {author} {\bibinfo {author} {\bibfnamefont {B.}~\bibnamefont
  {Salehian}}, \bibinfo {author} {\bibfnamefont {M.~A.}\ \bibnamefont {Gorji}},
  \bibinfo {author} {\bibfnamefont {H.}~\bibnamefont {Firouzjahi}},\ and\
  \bibinfo {author} {\bibfnamefont {S.}~\bibnamefont {Mukohyama}},\ }\bibfield
  {title} {\bibinfo {title} {{Vector dark matter production from inflation with
  symmetry breaking}},\ }\href {https://doi.org/10.1103/PhysRevD.103.063526}
  {\bibfield  {journal} {\bibinfo  {journal} {Phys. Rev. D}\ }\textbf {\bibinfo
  {volume} {103}},\ \bibinfo {pages} {063526} (\bibinfo {year}
  {2021}{\natexlab{a}})},\ \Eprint {https://arxiv.org/abs/2010.04491}
  {arXiv:2010.04491 [hep-ph]} \BibitemShut {NoStop}%
\bibitem [{\citenamefont {Nelson}\ and\ \citenamefont
  {Scholtz}(2011)}]{Nelson:2011sf}%
  \BibitemOpen
  \bibfield  {author} {\bibinfo {author} {\bibfnamefont {A.~E.}\ \bibnamefont
  {Nelson}}\ and\ \bibinfo {author} {\bibfnamefont {J.}~\bibnamefont
  {Scholtz}},\ }\bibfield  {title} {\bibinfo {title} {{Dark Light, Dark Matter
  and the Misalignment Mechanism}},\ }\href
  {https://doi.org/10.1103/PhysRevD.84.103501} {\bibfield  {journal} {\bibinfo
  {journal} {Phys. Rev. D}\ }\textbf {\bibinfo {volume} {84}},\ \bibinfo
  {pages} {103501} (\bibinfo {year} {2011})},\ \Eprint
  {https://arxiv.org/abs/1105.2812} {arXiv:1105.2812 [hep-ph]} \BibitemShut
  {NoStop}%
\bibitem [{\citenamefont {Arias}\ \emph {et~al.}(2012)\citenamefont {Arias},
  \citenamefont {Cadamuro}, \citenamefont {Goodsell}, \citenamefont {Jaeckel},
  \citenamefont {Redondo},\ and\ \citenamefont {Ringwald}}]{Arias:2012az}%
  \BibitemOpen
  \bibfield  {author} {\bibinfo {author} {\bibfnamefont {P.}~\bibnamefont
  {Arias}}, \bibinfo {author} {\bibfnamefont {D.}~\bibnamefont {Cadamuro}},
  \bibinfo {author} {\bibfnamefont {M.}~\bibnamefont {Goodsell}}, \bibinfo
  {author} {\bibfnamefont {J.}~\bibnamefont {Jaeckel}}, \bibinfo {author}
  {\bibfnamefont {J.}~\bibnamefont {Redondo}},\ and\ \bibinfo {author}
  {\bibfnamefont {A.}~\bibnamefont {Ringwald}},\ }\bibfield  {title} {\bibinfo
  {title} {{WISPy Cold Dark Matter}},\ }\href
  {https://doi.org/10.1088/1475-7516/2012/06/013} {\bibfield  {journal}
  {\bibinfo  {journal} {JCAP}\ }\textbf {\bibinfo {volume} {06}},\ \bibinfo
  {pages} {013}},\ \Eprint {https://arxiv.org/abs/1201.5902} {arXiv:1201.5902
  [hep-ph]} \BibitemShut {NoStop}%
\bibitem [{\citenamefont {Nakayama}(2019)}]{Nakayama:2019rhg}%
  \BibitemOpen
  \bibfield  {author} {\bibinfo {author} {\bibfnamefont {K.}~\bibnamefont
  {Nakayama}},\ }\bibfield  {title} {\bibinfo {title} {{Vector Coherent
  Oscillation Dark Matter}},\ }\href
  {https://doi.org/10.1088/1475-7516/2019/10/019} {\bibfield  {journal}
  {\bibinfo  {journal} {JCAP}\ }\textbf {\bibinfo {volume} {10}},\ \bibinfo
  {pages} {019}},\ \Eprint {https://arxiv.org/abs/1907.06243} {arXiv:1907.06243
  [hep-ph]} \BibitemShut {NoStop}%
\bibitem [{\citenamefont {Kitajima}\ and\ \citenamefont
  {Nakayama}(2023)}]{Kitajima:2023fun}%
  \BibitemOpen
  \bibfield  {author} {\bibinfo {author} {\bibfnamefont {N.}~\bibnamefont
  {Kitajima}}\ and\ \bibinfo {author} {\bibfnamefont {K.}~\bibnamefont
  {Nakayama}},\ }\bibfield  {title} {\bibinfo {title} {{Viable vector coherent
  oscillation dark~matter}},\ }\href
  {https://doi.org/10.1088/1475-7516/2023/07/014} {\bibfield  {journal}
  {\bibinfo  {journal} {JCAP}\ }\textbf {\bibinfo {volume} {07}},\ \bibinfo
  {pages} {014}},\ \Eprint {https://arxiv.org/abs/2303.04287} {arXiv:2303.04287
  [hep-ph]} \BibitemShut {NoStop}%
\bibitem [{\citenamefont {Hui}\ \emph {et~al.}(2017)\citenamefont {Hui},
  \citenamefont {Ostriker}, \citenamefont {Tremaine},\ and\ \citenamefont
  {Witten}}]{Hui:2016ltb}%
  \BibitemOpen
  \bibfield  {author} {\bibinfo {author} {\bibfnamefont {L.}~\bibnamefont
  {Hui}}, \bibinfo {author} {\bibfnamefont {J.~P.}\ \bibnamefont {Ostriker}},
  \bibinfo {author} {\bibfnamefont {S.}~\bibnamefont {Tremaine}},\ and\
  \bibinfo {author} {\bibfnamefont {E.}~\bibnamefont {Witten}},\ }\bibfield
  {title} {\bibinfo {title} {{Ultralight scalars as cosmological dark
  matter}},\ }\href {https://doi.org/10.1103/PhysRevD.95.043541} {\bibfield
  {journal} {\bibinfo  {journal} {Phys. Rev. D}\ }\textbf {\bibinfo {volume}
  {95}},\ \bibinfo {pages} {043541} (\bibinfo {year} {2017})},\ \Eprint
  {https://arxiv.org/abs/1610.08297} {arXiv:1610.08297 [astro-ph.CO]}
  \BibitemShut {NoStop}%
\bibitem [{\citenamefont {Hu}\ \emph {et~al.}(2000)\citenamefont {Hu},
  \citenamefont {Barkana},\ and\ \citenamefont {Gruzinov}}]{Hu:2000ke}%
  \BibitemOpen
  \bibfield  {author} {\bibinfo {author} {\bibfnamefont {W.}~\bibnamefont
  {Hu}}, \bibinfo {author} {\bibfnamefont {R.}~\bibnamefont {Barkana}},\ and\
  \bibinfo {author} {\bibfnamefont {A.}~\bibnamefont {Gruzinov}},\ }\bibfield
  {title} {\bibinfo {title} {{Cold and fuzzy dark matter}},\ }\href
  {https://doi.org/10.1103/PhysRevLett.85.1158} {\bibfield  {journal} {\bibinfo
   {journal} {Phys. Rev. Lett.}\ }\textbf {\bibinfo {volume} {85}},\ \bibinfo
  {pages} {1158} (\bibinfo {year} {2000})},\ \Eprint
  {https://arxiv.org/abs/astro-ph/0003365} {arXiv:astro-ph/0003365}
  \BibitemShut {NoStop}%
\bibitem [{\citenamefont {Ferreira}(2021)}]{Ferreira:2020fam}%
  \BibitemOpen
  \bibfield  {author} {\bibinfo {author} {\bibfnamefont {E.~G.~M.}\
  \bibnamefont {Ferreira}},\ }\bibfield  {title} {\bibinfo {title}
  {{Ultra-light dark matter}},\ }\href
  {https://doi.org/10.1007/s00159-021-00135-6} {\bibfield  {journal} {\bibinfo
  {journal} {Astron. Astrophys. Rev.}\ }\textbf {\bibinfo {volume} {29}},\
  \bibinfo {pages} {7} (\bibinfo {year} {2021})},\ \Eprint
  {https://arxiv.org/abs/2005.03254} {arXiv:2005.03254 [astro-ph.CO]}
  \BibitemShut {NoStop}%
\bibitem [{\citenamefont {Hui}(2021)}]{Hui:2021tkt}%
  \BibitemOpen
  \bibfield  {author} {\bibinfo {author} {\bibfnamefont {L.}~\bibnamefont
  {Hui}},\ }\bibfield  {title} {\bibinfo {title} {{Wave Dark Matter}},\ }\href
  {https://doi.org/10.1146/annurev-astro-120920-010024} {\bibfield  {journal}
  {\bibinfo  {journal} {Ann. Rev. Astron. Astrophys.}\ }\textbf {\bibinfo
  {volume} {59}},\ \bibinfo {pages} {247} (\bibinfo {year} {2021})},\ \Eprint
  {https://arxiv.org/abs/2101.11735} {arXiv:2101.11735 [astro-ph.CO]}
  \BibitemShut {NoStop}%
\bibitem [{\citenamefont {Salehian}\ \emph
  {et~al.}(2021{\natexlab{b}})\citenamefont {Salehian}, \citenamefont {Zhang},
  \citenamefont {Amin}, \citenamefont {Kaiser},\ and\ \citenamefont
  {Namjoo}}]{Salehian:2021khb}%
  \BibitemOpen
  \bibfield  {author} {\bibinfo {author} {\bibfnamefont {B.}~\bibnamefont
  {Salehian}}, \bibinfo {author} {\bibfnamefont {H.-Y.}\ \bibnamefont {Zhang}},
  \bibinfo {author} {\bibfnamefont {M.~A.}\ \bibnamefont {Amin}}, \bibinfo
  {author} {\bibfnamefont {D.~I.}\ \bibnamefont {Kaiser}},\ and\ \bibinfo
  {author} {\bibfnamefont {M.~H.}\ \bibnamefont {Namjoo}},\ }\bibfield  {title}
  {\bibinfo {title} {{Beyond Schr{\"o}dinger-Poisson: nonrelativistic effective
  field theory for scalar dark matter}},\ }\href
  {https://doi.org/10.1007/JHEP09(2021)050} {\bibfield  {journal} {\bibinfo
  {journal} {JHEP}\ }\textbf {\bibinfo {volume} {09}},\ \bibinfo {pages}
  {050}},\ \Eprint {https://arxiv.org/abs/2104.10128} {arXiv:2104.10128
  [astro-ph.CO]} \BibitemShut {NoStop}%
\bibitem [{\citenamefont {Salehian}\ \emph {et~al.}(2020)\citenamefont
  {Salehian}, \citenamefont {Namjoo},\ and\ \citenamefont
  {Kaiser}}]{Salehian:2020bon}%
  \BibitemOpen
  \bibfield  {author} {\bibinfo {author} {\bibfnamefont {B.}~\bibnamefont
  {Salehian}}, \bibinfo {author} {\bibfnamefont {M.~H.}\ \bibnamefont
  {Namjoo}},\ and\ \bibinfo {author} {\bibfnamefont {D.~I.}\ \bibnamefont
  {Kaiser}},\ }\bibfield  {title} {\bibinfo {title} {{Effective theories for a
  nonrelativistic field in an expanding universe: Induced self-interaction,
  pressure, sound speed, and viscosity}},\ }\href
  {https://doi.org/10.1007/JHEP07(2020)059} {\bibfield  {journal} {\bibinfo
  {journal} {JHEP}\ }\textbf {\bibinfo {volume} {07}},\ \bibinfo {pages}
  {059}},\ \Eprint {https://arxiv.org/abs/2005.05388} {arXiv:2005.05388
  [astro-ph.CO]} \BibitemShut {NoStop}%
\bibitem [{\citenamefont {Namjoo}\ \emph {et~al.}(2018)\citenamefont {Namjoo},
  \citenamefont {Guth},\ and\ \citenamefont {Kaiser}}]{Namjoo:2017nia}%
  \BibitemOpen
  \bibfield  {author} {\bibinfo {author} {\bibfnamefont {M.~H.}\ \bibnamefont
  {Namjoo}}, \bibinfo {author} {\bibfnamefont {A.~H.}\ \bibnamefont {Guth}},\
  and\ \bibinfo {author} {\bibfnamefont {D.~I.}\ \bibnamefont {Kaiser}},\
  }\bibfield  {title} {\bibinfo {title} {{Relativistic Corrections to
  Nonrelativistic Effective Field Theories}},\ }\href
  {https://doi.org/10.1103/PhysRevD.98.016011} {\bibfield  {journal} {\bibinfo
  {journal} {Phys. Rev. D}\ }\textbf {\bibinfo {volume} {98}},\ \bibinfo
  {pages} {016011} (\bibinfo {year} {2018})},\ \Eprint
  {https://arxiv.org/abs/1712.00445} {arXiv:1712.00445 [hep-ph]} \BibitemShut
  {NoStop}%
\bibitem [{\citenamefont {Braaten}\ \emph {et~al.}(2018)\citenamefont
  {Braaten}, \citenamefont {Mohapatra},\ and\ \citenamefont
  {Zhang}}]{Braaten:2018lmj}%
  \BibitemOpen
  \bibfield  {author} {\bibinfo {author} {\bibfnamefont {E.}~\bibnamefont
  {Braaten}}, \bibinfo {author} {\bibfnamefont {A.}~\bibnamefont {Mohapatra}},\
  and\ \bibinfo {author} {\bibfnamefont {H.}~\bibnamefont {Zhang}},\ }\bibfield
   {title} {\bibinfo {title} {{Classical Nonrelativistic Effective Field
  Theories for a Real Scalar Field}},\ }\href
  {https://doi.org/10.1103/PhysRevD.98.096012} {\bibfield  {journal} {\bibinfo
  {journal} {Phys. Rev. D}\ }\textbf {\bibinfo {volume} {98}},\ \bibinfo
  {pages} {096012} (\bibinfo {year} {2018})},\ \Eprint
  {https://arxiv.org/abs/1806.01898} {arXiv:1806.01898 [hep-ph]} \BibitemShut
  {NoStop}%
\bibitem [{\citenamefont {Schive}\ \emph
  {et~al.}(2014{\natexlab{a}})\citenamefont {Schive}, \citenamefont {Chiueh},\
  and\ \citenamefont {Broadhurst}}]{Schive:2014dra}%
  \BibitemOpen
  \bibfield  {author} {\bibinfo {author} {\bibfnamefont {H.-Y.}\ \bibnamefont
  {Schive}}, \bibinfo {author} {\bibfnamefont {T.}~\bibnamefont {Chiueh}},\
  and\ \bibinfo {author} {\bibfnamefont {T.}~\bibnamefont {Broadhurst}},\
  }\bibfield  {title} {\bibinfo {title} {{Cosmic Structure as the Quantum
  Interference of a Coherent Dark Wave}},\ }\href
  {https://doi.org/10.1038/nphys2996} {\bibfield  {journal} {\bibinfo
  {journal} {Nature Phys.}\ }\textbf {\bibinfo {volume} {10}},\ \bibinfo
  {pages} {496} (\bibinfo {year} {2014}{\natexlab{a}})},\ \Eprint
  {https://arxiv.org/abs/1406.6586} {arXiv:1406.6586 [astro-ph.GA]}
  \BibitemShut {NoStop}%
\bibitem [{\citenamefont {Schive}\ \emph
  {et~al.}(2014{\natexlab{b}})\citenamefont {Schive}, \citenamefont {Liao},
  \citenamefont {Woo}, \citenamefont {Wong}, \citenamefont {Chiueh},
  \citenamefont {Broadhurst},\ and\ \citenamefont {Hwang}}]{Schive:2014hza}%
  \BibitemOpen
  \bibfield  {author} {\bibinfo {author} {\bibfnamefont {H.-Y.}\ \bibnamefont
  {Schive}}, \bibinfo {author} {\bibfnamefont {M.-H.}\ \bibnamefont {Liao}},
  \bibinfo {author} {\bibfnamefont {T.-P.}\ \bibnamefont {Woo}}, \bibinfo
  {author} {\bibfnamefont {S.-K.}\ \bibnamefont {Wong}}, \bibinfo {author}
  {\bibfnamefont {T.}~\bibnamefont {Chiueh}}, \bibinfo {author} {\bibfnamefont
  {T.}~\bibnamefont {Broadhurst}},\ and\ \bibinfo {author} {\bibfnamefont
  {W.~Y.~P.}\ \bibnamefont {Hwang}},\ }\bibfield  {title} {\bibinfo {title}
  {{Understanding the Core-Halo Relation of Quantum Wave Dark Matter from 3D
  Simulations}},\ }\href {https://doi.org/10.1103/PhysRevLett.113.261302}
  {\bibfield  {journal} {\bibinfo  {journal} {Phys. Rev. Lett.}\ }\textbf
  {\bibinfo {volume} {113}},\ \bibinfo {pages} {261302} (\bibinfo {year}
  {2014}{\natexlab{b}})},\ \Eprint {https://arxiv.org/abs/1407.7762}
  {arXiv:1407.7762 [astro-ph.GA]} \BibitemShut {NoStop}%
\bibitem [{\citenamefont {Veltmaat}\ and\ \citenamefont
  {Niemeyer}(2016)}]{Veltmaat:2016rxo}%
  \BibitemOpen
  \bibfield  {author} {\bibinfo {author} {\bibfnamefont {J.}~\bibnamefont
  {Veltmaat}}\ and\ \bibinfo {author} {\bibfnamefont {J.~C.}\ \bibnamefont
  {Niemeyer}},\ }\bibfield  {title} {\bibinfo {title} {{Cosmological
  particle-in-cell simulations with ultralight axion dark matter}},\ }\href
  {https://doi.org/10.1103/PhysRevD.94.123523} {\bibfield  {journal} {\bibinfo
  {journal} {Phys. Rev. D}\ }\textbf {\bibinfo {volume} {94}},\ \bibinfo
  {pages} {123523} (\bibinfo {year} {2016})},\ \Eprint
  {https://arxiv.org/abs/1608.00802} {arXiv:1608.00802 [astro-ph.CO]}
  \BibitemShut {NoStop}%
\bibitem [{\citenamefont {Edwards}\ \emph {et~al.}(2018)\citenamefont
  {Edwards}, \citenamefont {Kendall}, \citenamefont {Hotchkiss},\ and\
  \citenamefont {Easther}}]{Edwards:2018ccc}%
  \BibitemOpen
  \bibfield  {author} {\bibinfo {author} {\bibfnamefont {F.}~\bibnamefont
  {Edwards}}, \bibinfo {author} {\bibfnamefont {E.}~\bibnamefont {Kendall}},
  \bibinfo {author} {\bibfnamefont {S.}~\bibnamefont {Hotchkiss}},\ and\
  \bibinfo {author} {\bibfnamefont {R.}~\bibnamefont {Easther}},\ }\bibfield
  {title} {\bibinfo {title} {{PyUltraLight: A Pseudo-Spectral Solver for
  Ultralight Dark Matter Dynamics}},\ }\href
  {https://doi.org/10.1088/1475-7516/2018/10/027} {\bibfield  {journal}
  {\bibinfo  {journal} {JCAP}\ }\textbf {\bibinfo {volume} {10}},\ \bibinfo
  {pages} {027}},\ \Eprint {https://arxiv.org/abs/1807.04037} {arXiv:1807.04037
  [astro-ph.CO]} \BibitemShut {NoStop}%
\bibitem [{\citenamefont {Li}\ \emph {et~al.}(2019)\citenamefont {Li},
  \citenamefont {Hui},\ and\ \citenamefont {Bryan}}]{Li:2018kyk}%
  \BibitemOpen
  \bibfield  {author} {\bibinfo {author} {\bibfnamefont {X.}~\bibnamefont
  {Li}}, \bibinfo {author} {\bibfnamefont {L.}~\bibnamefont {Hui}},\ and\
  \bibinfo {author} {\bibfnamefont {G.~L.}\ \bibnamefont {Bryan}},\ }\bibfield
  {title} {\bibinfo {title} {{Numerical and Perturbative Computations of the
  Fuzzy Dark Matter Model}},\ }\href
  {https://doi.org/10.1103/PhysRevD.99.063509} {\bibfield  {journal} {\bibinfo
  {journal} {Phys. Rev. D}\ }\textbf {\bibinfo {volume} {99}},\ \bibinfo
  {pages} {063509} (\bibinfo {year} {2019})},\ \Eprint
  {https://arxiv.org/abs/1810.01915} {arXiv:1810.01915 [astro-ph.CO]}
  \BibitemShut {NoStop}%
\bibitem [{\citenamefont {Bar}\ \emph {et~al.}(2018)\citenamefont {Bar},
  \citenamefont {Blas}, \citenamefont {Blum},\ and\ \citenamefont
  {Sibiryakov}}]{Bar:2018acw}%
  \BibitemOpen
  \bibfield  {author} {\bibinfo {author} {\bibfnamefont {N.}~\bibnamefont
  {Bar}}, \bibinfo {author} {\bibfnamefont {D.}~\bibnamefont {Blas}}, \bibinfo
  {author} {\bibfnamefont {K.}~\bibnamefont {Blum}},\ and\ \bibinfo {author}
  {\bibfnamefont {S.}~\bibnamefont {Sibiryakov}},\ }\bibfield  {title}
  {\bibinfo {title} {{Galactic rotation curves versus ultralight dark matter:
  Implications of the soliton-host halo relation}},\ }\href
  {https://doi.org/10.1103/PhysRevD.98.083027} {\bibfield  {journal} {\bibinfo
  {journal} {Phys. Rev. D}\ }\textbf {\bibinfo {volume} {98}},\ \bibinfo
  {pages} {083027} (\bibinfo {year} {2018})},\ \Eprint
  {https://arxiv.org/abs/1805.00122} {arXiv:1805.00122 [astro-ph.CO]}
  \BibitemShut {NoStop}%
\bibitem [{\citenamefont {Li}\ \emph {et~al.}(2021)\citenamefont {Li},
  \citenamefont {Hui},\ and\ \citenamefont {Yavetz}}]{Li:2020ryg}%
  \BibitemOpen
  \bibfield  {author} {\bibinfo {author} {\bibfnamefont {X.}~\bibnamefont
  {Li}}, \bibinfo {author} {\bibfnamefont {L.}~\bibnamefont {Hui}},\ and\
  \bibinfo {author} {\bibfnamefont {T.~D.}\ \bibnamefont {Yavetz}},\ }\bibfield
   {title} {\bibinfo {title} {{Oscillations and Random Walk of the Soliton Core
  in a Fuzzy Dark Matter Halo}},\ }\href
  {https://doi.org/10.1103/PhysRevD.103.023508} {\bibfield  {journal} {\bibinfo
   {journal} {Phys. Rev. D}\ }\textbf {\bibinfo {volume} {103}},\ \bibinfo
  {pages} {023508} (\bibinfo {year} {2021})},\ \Eprint
  {https://arxiv.org/abs/2011.11416} {arXiv:2011.11416 [astro-ph.CO]}
  \BibitemShut {NoStop}%
\bibitem [{\citenamefont {Chen}\ \emph {et~al.}(2021)\citenamefont {Chen},
  \citenamefont {Du}, \citenamefont {Lentz}, \citenamefont {Marsh},\ and\
  \citenamefont {Niemeyer}}]{Chen:2020cef}%
  \BibitemOpen
  \bibfield  {author} {\bibinfo {author} {\bibfnamefont {J.}~\bibnamefont
  {Chen}}, \bibinfo {author} {\bibfnamefont {X.}~\bibnamefont {Du}}, \bibinfo
  {author} {\bibfnamefont {E.~W.}\ \bibnamefont {Lentz}}, \bibinfo {author}
  {\bibfnamefont {D.~J.~E.}\ \bibnamefont {Marsh}},\ and\ \bibinfo {author}
  {\bibfnamefont {J.~C.}\ \bibnamefont {Niemeyer}},\ }\bibfield  {title}
  {\bibinfo {title} {{New insights into the formation and growth of boson stars
  in dark matter halos}},\ }\href {https://doi.org/10.1103/PhysRevD.104.083022}
  {\bibfield  {journal} {\bibinfo  {journal} {Phys. Rev. D}\ }\textbf {\bibinfo
  {volume} {104}},\ \bibinfo {pages} {083022} (\bibinfo {year} {2021})},\
  \Eprint {https://arxiv.org/abs/2011.01333} {arXiv:2011.01333 [astro-ph.CO]}
  \BibitemShut {NoStop}%
\bibitem [{\citenamefont {May}\ and\ \citenamefont
  {Springel}(2021)}]{May:2021wwp}%
  \BibitemOpen
  \bibfield  {author} {\bibinfo {author} {\bibfnamefont {S.}~\bibnamefont
  {May}}\ and\ \bibinfo {author} {\bibfnamefont {V.}~\bibnamefont {Springel}},\
  }\bibfield  {title} {\bibinfo {title} {{Structure formation in large-volume
  cosmological simulations of fuzzy dark matter: impact of the non-linear
  dynamics}},\ }\href {https://doi.org/10.1093/mnras/stab1764} {\bibfield
  {journal} {\bibinfo  {journal} {Mon. Not. Roy. Astron. Soc.}\ }\textbf
  {\bibinfo {volume} {506}},\ \bibinfo {pages} {2603} (\bibinfo {year}
  {2021})},\ \Eprint {https://arxiv.org/abs/2101.01828} {arXiv:2101.01828
  [astro-ph.CO]} \BibitemShut {NoStop}%
\bibitem [{\citenamefont {Chan}\ \emph {et~al.}(2022)\citenamefont {Chan},
  \citenamefont {Ferreira}, \citenamefont {May}, \citenamefont {Hayashi},\ and\
  \citenamefont {Chiba}}]{Chan:2021bja}%
  \BibitemOpen
  \bibfield  {author} {\bibinfo {author} {\bibfnamefont {H.~Y.~J.}\
  \bibnamefont {Chan}}, \bibinfo {author} {\bibfnamefont {E.~G.~M.}\
  \bibnamefont {Ferreira}}, \bibinfo {author} {\bibfnamefont {S.}~\bibnamefont
  {May}}, \bibinfo {author} {\bibfnamefont {K.}~\bibnamefont {Hayashi}},\ and\
  \bibinfo {author} {\bibfnamefont {M.}~\bibnamefont {Chiba}},\ }\bibfield
  {title} {\bibinfo {title} {{The diversity of core{\textendash}halo structure
  in the fuzzy dark matter model}},\ }\href
  {https://doi.org/10.1093/mnras/stac063} {\bibfield  {journal} {\bibinfo
  {journal} {Mon. Not. Roy. Astron. Soc.}\ }\textbf {\bibinfo {volume} {511}},\
  \bibinfo {pages} {943} (\bibinfo {year} {2022})},\ \Eprint
  {https://arxiv.org/abs/2110.11882} {arXiv:2110.11882 [astro-ph.CO]}
  \BibitemShut {NoStop}%
\bibitem [{\citenamefont {Jain}\ and\ \citenamefont
  {Amin}(2023)}]{Jain:2022agt}%
  \BibitemOpen
  \bibfield  {author} {\bibinfo {author} {\bibfnamefont {M.}~\bibnamefont
  {Jain}}\ and\ \bibinfo {author} {\bibfnamefont {M.~A.}\ \bibnamefont
  {Amin}},\ }\bibfield  {title} {\bibinfo {title} {{i-SPin: an integrator for
  multicomponent Schr{\"o}dinger-Poisson systems with self-interactions}},\
  }\href {https://doi.org/10.1088/1475-7516/2023/04/053} {\bibfield  {journal}
  {\bibinfo  {journal} {JCAP}\ }\textbf {\bibinfo {volume} {04}},\ \bibinfo
  {pages} {053}},\ \Eprint {https://arxiv.org/abs/2211.08433} {arXiv:2211.08433
  [astro-ph.CO]} \BibitemShut {NoStop}%
\bibitem [{\citenamefont {Amin}\ \emph {et~al.}(2022)\citenamefont {Amin},
  \citenamefont {Jain}, \citenamefont {Karur},\ and\ \citenamefont
  {Mocz}}]{Amin:2022pzv}%
  \BibitemOpen
  \bibfield  {author} {\bibinfo {author} {\bibfnamefont {M.~A.}\ \bibnamefont
  {Amin}}, \bibinfo {author} {\bibfnamefont {M.}~\bibnamefont {Jain}}, \bibinfo
  {author} {\bibfnamefont {R.}~\bibnamefont {Karur}},\ and\ \bibinfo {author}
  {\bibfnamefont {P.}~\bibnamefont {Mocz}},\ }\bibfield  {title} {\bibinfo
  {title} {{Small-scale structure in vector dark matter}},\ }\href
  {https://doi.org/10.1088/1475-7516/2022/08/014} {\bibfield  {journal}
  {\bibinfo  {journal} {JCAP}\ }\textbf {\bibinfo {volume} {08}}\bibfield
  {number} {\bibinfo  {number} { (08)},\ \bibinfo {pages} {014}},\ }\Eprint
  {https://arxiv.org/abs/2203.11935} {arXiv:2203.11935 [astro-ph.CO]}
  \BibitemShut {NoStop}%
\bibitem [{\citenamefont {Jain}\ \emph {et~al.}(2024)\citenamefont {Jain},
  \citenamefont {Wanichwecharungruang},\ and\ \citenamefont
  {Thomas}}]{Jain:2023tsr}%
  \BibitemOpen
  \bibfield  {author} {\bibinfo {author} {\bibfnamefont {M.}~\bibnamefont
  {Jain}}, \bibinfo {author} {\bibfnamefont {W.}~\bibnamefont
  {Wanichwecharungruang}},\ and\ \bibinfo {author} {\bibfnamefont
  {J.}~\bibnamefont {Thomas}},\ }\bibfield  {title} {\bibinfo {title} {{Kinetic
  relaxation and nucleation of Bose stars in self-interacting wave dark
  matter}},\ }\href {https://doi.org/10.1103/PhysRevD.109.016002} {\bibfield
  {journal} {\bibinfo  {journal} {Phys. Rev. D}\ }\textbf {\bibinfo {volume}
  {109}},\ \bibinfo {pages} {016002} (\bibinfo {year} {2024})},\ \Eprint
  {https://arxiv.org/abs/2310.00058} {arXiv:2310.00058 [astro-ph.CO]}
  \BibitemShut {NoStop}%
\bibitem [{\citenamefont {Nambo}\ \emph {et~al.}(2025)\citenamefont {Nambo},
  \citenamefont {Diez-Tejedor}, \citenamefont {Preciado-Govea}, \citenamefont
  {Roque},\ and\ \citenamefont {Sarbach}}]{Nambo:2024hao}%
  \BibitemOpen
  \bibfield  {author} {\bibinfo {author} {\bibfnamefont {E.~C.}\ \bibnamefont
  {Nambo}}, \bibinfo {author} {\bibfnamefont {A.}~\bibnamefont {Diez-Tejedor}},
  \bibinfo {author} {\bibfnamefont {E.}~\bibnamefont {Preciado-Govea}},
  \bibinfo {author} {\bibfnamefont {A.~A.}\ \bibnamefont {Roque}},\ and\
  \bibinfo {author} {\bibfnamefont {O.}~\bibnamefont {Sarbach}},\ }\bibfield
  {title} {\bibinfo {title} {{Nonrelativistic Proca stars: Spherical stationary
  and multifrequency states}},\ }\href
  {https://doi.org/10.1103/PhysRevD.111.064065} {\bibfield  {journal} {\bibinfo
   {journal} {Phys. Rev. D}\ }\textbf {\bibinfo {volume} {111}},\ \bibinfo
  {pages} {064065} (\bibinfo {year} {2025})},\ \Eprint
  {https://arxiv.org/abs/2412.06901} {arXiv:2412.06901 [gr-qc]} \BibitemShut
  {NoStop}%
\bibitem [{\citenamefont {Hu}\ and\ \citenamefont
  {Sugiyama}(1996)}]{Hu:1995en}%
  \BibitemOpen
  \bibfield  {author} {\bibinfo {author} {\bibfnamefont {W.}~\bibnamefont
  {Hu}}\ and\ \bibinfo {author} {\bibfnamefont {N.}~\bibnamefont {Sugiyama}},\
  }\bibfield  {title} {\bibinfo {title} {{Small scale cosmological
  perturbations: An Analytic approach}},\ }\href
  {https://doi.org/10.1086/177989} {\bibfield  {journal} {\bibinfo  {journal}
  {Astrophys. J.}\ }\textbf {\bibinfo {volume} {471}},\ \bibinfo {pages} {542}
  (\bibinfo {year} {1996})},\ \Eprint {https://arxiv.org/abs/astro-ph/9510117}
  {arXiv:astro-ph/9510117} \BibitemShut {NoStop}%
\bibitem [{\citenamefont {Baumann}(2022)}]{Baumann:2022mni}%
  \BibitemOpen
  \bibfield  {author} {\bibinfo {author} {\bibfnamefont {D.}~\bibnamefont
  {Baumann}},\ }\href {https://doi.org/10.1017/9781108937092} {\emph {\bibinfo
  {title} {{Cosmology}}}}\ (\bibinfo  {publisher} {Cambridge University
  Press},\ \bibinfo {year} {2022})\BibitemShut {NoStop}%
\bibitem [{\citenamefont {Amin}\ and\ \citenamefont
  {Mirbabayi}(2024)}]{Amin:2022nlh}%
  \BibitemOpen
  \bibfield  {author} {\bibinfo {author} {\bibfnamefont {M.~A.}\ \bibnamefont
  {Amin}}\ and\ \bibinfo {author} {\bibfnamefont {M.}~\bibnamefont
  {Mirbabayi}},\ }\bibfield  {title} {\bibinfo {title} {{A Lower Bound on Dark
  Matter Mass}},\ }\href {https://doi.org/10.1103/PhysRevLett.132.221004}
  {\bibfield  {journal} {\bibinfo  {journal} {Phys. Rev. Lett.}\ }\textbf
  {\bibinfo {volume} {132}},\ \bibinfo {pages} {221004} (\bibinfo {year}
  {2024})},\ \Eprint {https://arxiv.org/abs/2211.09775} {arXiv:2211.09775
  [hep-ph]} \BibitemShut {NoStop}%
\bibitem [{\citenamefont {Liu}\ \emph {et~al.}(2025{\natexlab{a}})\citenamefont
  {Liu}, \citenamefont {Hu},\ and\ \citenamefont {Xiao}}]{Liu:2024pjg}%
  \BibitemOpen
  \bibfield  {author} {\bibinfo {author} {\bibfnamefont {R.}~\bibnamefont
  {Liu}}, \bibinfo {author} {\bibfnamefont {W.}~\bibnamefont {Hu}},\ and\
  \bibinfo {author} {\bibfnamefont {H.}~\bibnamefont {Xiao}},\ }\bibfield
  {title} {\bibinfo {title} {{Warm and fuzzy dark matter: Free streaming of
  wave dark matter}},\ }\href {https://doi.org/10.1103/PhysRevD.111.023535}
  {\bibfield  {journal} {\bibinfo  {journal} {Phys. Rev. D}\ }\textbf {\bibinfo
  {volume} {111}},\ \bibinfo {pages} {023535} (\bibinfo {year}
  {2025}{\natexlab{a}})},\ \Eprint {https://arxiv.org/abs/2406.12970}
  {arXiv:2406.12970 [hep-ph]} \BibitemShut {NoStop}%
\bibitem [{\citenamefont {Liu}\ \emph {et~al.}(2025{\natexlab{b}})\citenamefont
  {Liu}, \citenamefont {Hu},\ and\ \citenamefont {Xiao}}]{Liu:2025lts}%
  \BibitemOpen
  \bibfield  {author} {\bibinfo {author} {\bibfnamefont {R.}~\bibnamefont
  {Liu}}, \bibinfo {author} {\bibfnamefont {W.}~\bibnamefont {Hu}},\ and\
  \bibinfo {author} {\bibfnamefont {H.}~\bibnamefont {Xiao}},\ }\bibfield
  {title} {\bibinfo {title} {{Interference with Gravitational Instability: Hot
  and Fuzzy Dark Matter}},\ }\href@noop {} {\  (\bibinfo {year}
  {2025}{\natexlab{b}})},\ \Eprint {https://arxiv.org/abs/2504.01937}
  {arXiv:2504.01937 [astro-ph.CO]} \BibitemShut {NoStop}%
\bibitem [{\citenamefont {Capanelli}\ \emph {et~al.}(2025)\citenamefont
  {Capanelli}, \citenamefont {Hu},\ and\ \citenamefont
  {McDonough}}]{Capanelli:2025nrj}%
  \BibitemOpen
  \bibfield  {author} {\bibinfo {author} {\bibfnamefont {C.}~\bibnamefont
  {Capanelli}}, \bibinfo {author} {\bibfnamefont {W.}~\bibnamefont {Hu}},\ and\
  \bibinfo {author} {\bibfnamefont {E.}~\bibnamefont {McDonough}},\ }\bibfield
  {title} {\bibinfo {title} {{Wave Interference in Self-Interacting Fuzzy Dark
  Matter}},\ }\href@noop {} {\  (\bibinfo {year} {2025})},\ \Eprint
  {https://arxiv.org/abs/2503.21865} {arXiv:2503.21865 [astro-ph.CO]}
  \BibitemShut {NoStop}%
\bibitem [{\citenamefont {Amin}\ \emph
  {et~al.}(2025{\natexlab{b}})\citenamefont {Amin}, \citenamefont {May},\ and\
  \citenamefont {Mirbabayi}}]{Amin:2025sla}%
  \BibitemOpen
  \bibfield  {author} {\bibinfo {author} {\bibfnamefont {M.~A.}\ \bibnamefont
  {Amin}}, \bibinfo {author} {\bibfnamefont {S.}~\bibnamefont {May}},\ and\
  \bibinfo {author} {\bibfnamefont {M.}~\bibnamefont {Mirbabayi}},\ }\bibfield
  {title} {\bibinfo {title} {{Early Growth of Structure in Warm Wave Dark
  Matter}},\ }\href@noop {} {\  (\bibinfo {year} {2025}{\natexlab{b}})},\
  \Eprint {https://arxiv.org/abs/2506.12131} {arXiv:2506.12131 [astro-ph.CO]}
  \BibitemShut {NoStop}%
\bibitem [{\citenamefont {Scoccimarro}\ \emph {et~al.}(2012)\citenamefont
  {Scoccimarro}, \citenamefont {Hui}, \citenamefont {Manera},\ and\
  \citenamefont {Chan}}]{Scoccimarro:2011pz}%
  \BibitemOpen
  \bibfield  {author} {\bibinfo {author} {\bibfnamefont {R.}~\bibnamefont
  {Scoccimarro}}, \bibinfo {author} {\bibfnamefont {L.}~\bibnamefont {Hui}},
  \bibinfo {author} {\bibfnamefont {M.}~\bibnamefont {Manera}},\ and\ \bibinfo
  {author} {\bibfnamefont {K.~C.}\ \bibnamefont {Chan}},\ }\bibfield  {title}
  {\bibinfo {title} {{Large-scale Bias and Efficient Generation of Initial
  Conditions for Non-Local Primordial Non-Gaussianity}},\ }\href
  {https://doi.org/10.1103/PhysRevD.85.083002} {\bibfield  {journal} {\bibinfo
  {journal} {Phys. Rev. D}\ }\textbf {\bibinfo {volume} {85}},\ \bibinfo
  {pages} {083002} (\bibinfo {year} {2012})},\ \Eprint
  {https://arxiv.org/abs/1108.5512} {arXiv:1108.5512 [astro-ph.CO]}
  \BibitemShut {NoStop}%
\bibitem [{\citenamefont {Coulton}\ \emph {et~al.}(2024)\citenamefont
  {Coulton}, \citenamefont {Philcox},\ and\ \citenamefont
  {Villaescusa-Navarro}}]{Coulton:2023oug}%
  \BibitemOpen
  \bibfield  {author} {\bibinfo {author} {\bibfnamefont {W.~R.}\ \bibnamefont
  {Coulton}}, \bibinfo {author} {\bibfnamefont {O.~H.~E.}\ \bibnamefont
  {Philcox}},\ and\ \bibinfo {author} {\bibfnamefont {F.}~\bibnamefont
  {Villaescusa-Navarro}},\ }\bibfield  {title} {\bibinfo {title} {{Signatures
  of a parity-violating universe}},\ }\href
  {https://doi.org/10.1103/PhysRevD.109.023531} {\bibfield  {journal} {\bibinfo
   {journal} {Phys. Rev. D}\ }\textbf {\bibinfo {volume} {109}},\ \bibinfo
  {pages} {023531} (\bibinfo {year} {2024})},\ \Eprint
  {https://arxiv.org/abs/2306.11782} {arXiv:2306.11782 [astro-ph.CO]}
  \BibitemShut {NoStop}%
\bibitem [{\citenamefont {Bao}\ \emph {et~al.}(2025)\citenamefont {Bao},
  \citenamefont {Wang}, \citenamefont {Xianyu},\ and\ \citenamefont
  {Zhong}}]{Bao:2025onc}%
  \BibitemOpen
  \bibfield  {author} {\bibinfo {author} {\bibfnamefont {Y.}~\bibnamefont
  {Bao}}, \bibinfo {author} {\bibfnamefont {L.-T.}\ \bibnamefont {Wang}},
  \bibinfo {author} {\bibfnamefont {Z.-Z.}\ \bibnamefont {Xianyu}},\ and\
  \bibinfo {author} {\bibfnamefont {Y.-M.}\ \bibnamefont {Zhong}},\ }\bibfield
  {title} {\bibinfo {title} {{Anatomy of Parity-violating Trispectra in Galaxy
  Surveys}},\ }\href@noop {} {\  (\bibinfo {year} {2025})},\ \Eprint
  {https://arxiv.org/abs/2504.02931} {arXiv:2504.02931 [astro-ph.CO]}
  \BibitemShut {NoStop}%
\end{thebibliography}%
\
\newpage
\includepdf[pages=1]{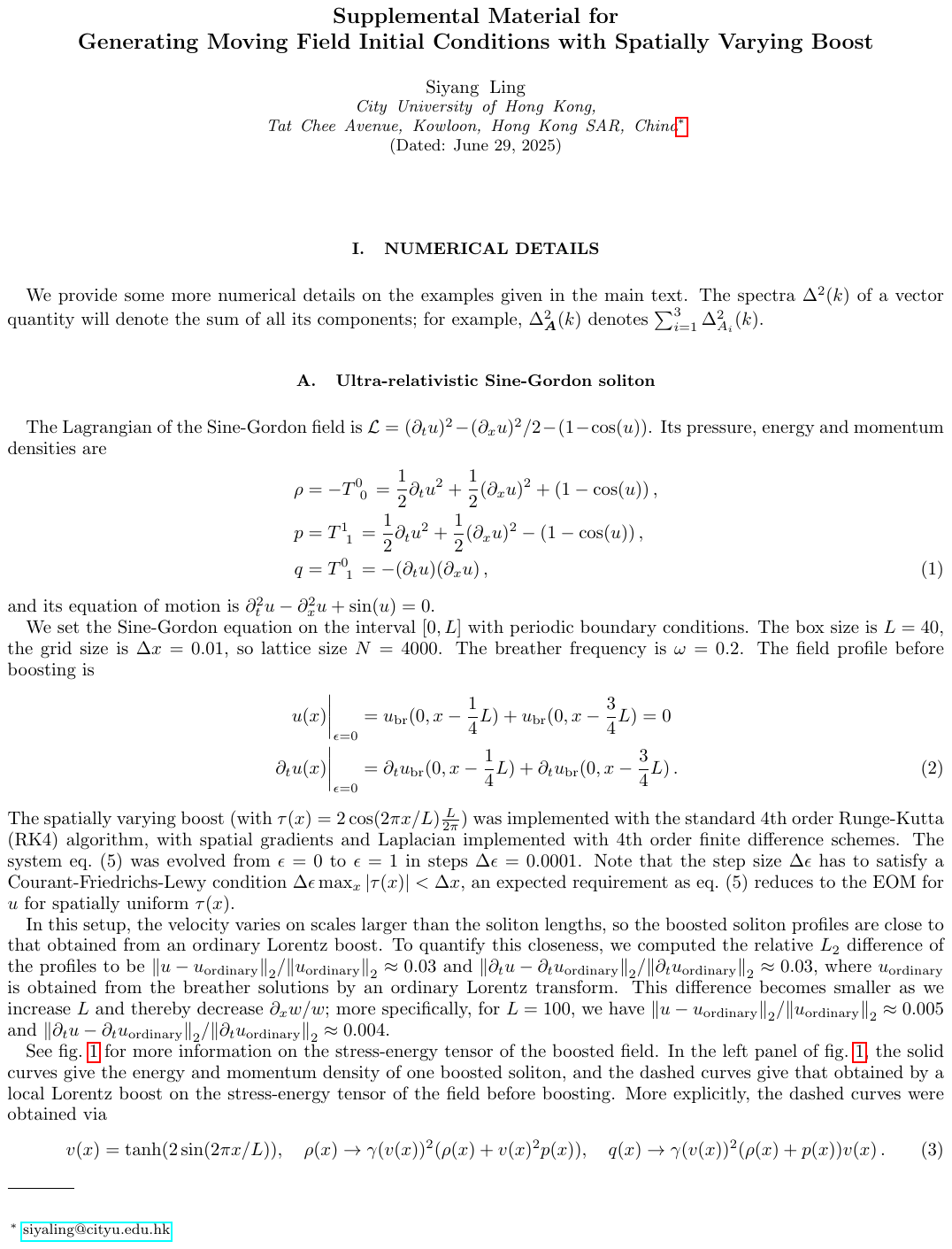}
\newpage
\
\newpage
\includepdf[pages=2]{supp.pdf}
\newpage
\
\newpage
\includepdf[pages=3]{supp.pdf}
\newpage
\
\newpage


\section*{Supplementary material}

# Supplemental Material for
# Generating Moving Field Initial Conditions with Spatially Varying Boost

Siyang Ling
*City University of Hong Kong,
Tat Chee Avenue, Kowloon, Hong Kong SAR, China*[*]
(Dated: June 29, 2025)

## I. NUMERICAL DETAILS

We provide some more numerical details on the examples given in the main text. The spectra $\Delta^2(k)$ of a vector quantity will denote the sum of all its components; for example, $\Delta^2_{\boldsymbol{A}}(k)$ denotes $\sum_{i=1}^{3}\Delta^2_{A_i}(k)$.

### A. Ultra-relativistic Sine-Gordon soliton

The Lagrangian of the Sine-Gordon field is $\mathcal{L} = (\partial_t u)^2 - (\partial_x u)^2 / 2 - (1 - \cos(u))$. Its pressure, energy and momentum densities are

$$\begin{aligned}
\rho = -T^0{}_0 &= \frac{1}{2}\partial_t u^2 + \frac{1}{2}(\partial_x u)^2 + (1 - \cos(u)),\\
p = T^1{}_1 &= \frac{1}{2}\partial_t u^2 + \frac{1}{2}(\partial_x u)^2 - (1 - \cos(u)),\\
q = T^0{}_1 &= -(\partial_t u)(\partial_x u),
\end{aligned} \quad (1)$$

and its equation of motion is $\partial_t^2 u - \partial_x^2 u + \sin(u) = 0$.

We set the Sine-Gordon equation on the interval $[0, L]$ with periodic boundary conditions. The box size is $L = 40$, the grid size is $\Delta x = 0.01$, so lattice size $N = 4000$. The breather frequency is $\omega = 0.2$. The field profile before boosting is

$$\begin{aligned}
u(x)\Big|_{\epsilon=0} &= u_{\mathrm{br}}(0, x - \tfrac{1}{4}L) + u_{\mathrm{br}}(0, x - \tfrac{3}{4}L) = 0\\
\partial_t u(x)\Big|_{\epsilon=0} &= \partial_t u_{\mathrm{br}}(0, x - \tfrac{1}{4}L) + \partial_t u_{\mathrm{br}}(0, x - \tfrac{3}{4}L).
\end{aligned} \quad (2)$$

The spatially varying boost (with $\tau(x) = 2\cos(2\pi x/L)\frac{L}{2\pi}$) was implemented with the standard 4th order Runge-Kutta (RK4) algorithm, with spatial gradients and Laplacian implemented with 4th order finite difference schemes. The system eq. (5) was evolved from $\epsilon = 0$ to $\epsilon = 1$ in steps $\Delta\epsilon = 0.0001$. Note that the step size $\Delta\epsilon$ has to satisfy a Courant-Friedrichs-Lewy condition $\Delta\epsilon \max_x |\tau(x)| < \Delta x$, an expected requirement as eq. (5) reduces to the EOM for $u$ for spatially uniform $\tau(x)$.

In this setup, the velocity varies on scales larger than the soliton lengths, so the boosted soliton profiles are close to that obtained from an ordinary Lorentz boost. To quantify this closeness, we computed the relative $L_2$ difference of the profiles to be $\|u - u_{\mathrm{ordinary}}\|_2 / \|u_{\mathrm{ordinary}}\|_2 \approx 0.03$ and $\|\partial_t u - \partial_t u_{\mathrm{ordinary}}\|_2 / \|\partial_t u_{\mathrm{ordinary}}\|_2 \approx 0.03$, where $u_{\mathrm{ordinary}}$ is obtained from the breather solutions by an ordinary Lorentz transform. This difference becomes smaller as we increase $L$ and thereby decrease $\partial_x w / w$; more specifically, for $L = 100$, we have $\|u - u_{\mathrm{ordinary}}\|_2 / \|u_{\mathrm{ordinary}}\|_2 \approx 0.005$ and $\|\partial_t u - \partial_t u_{\mathrm{ordinary}}\|_2 / \|\partial_t u_{\mathrm{ordinary}}\|_2 \approx 0.004$.

See fig. 1 for more information on the stress-energy tensor of the boosted field. In the left panel of fig. 1, the solid curves give the energy and momentum density of one boosted soliton, and the dashed curves give that obtained by a local Lorentz boost on the stress-energy tensor of the field before boosting. More explicitly, the dashed curves were obtained via

$$v(x) = \tanh(2\sin(2\pi x/L)), \quad \rho(x) \to \gamma(v(x))^2(\rho(x) + v(x)^2 p(x)), \quad q(x) \to \gamma(v(x))^2(\rho(x) + p(x))v(x). \quad (3)$$

---

[*] siyaling@cityu.edu.hk

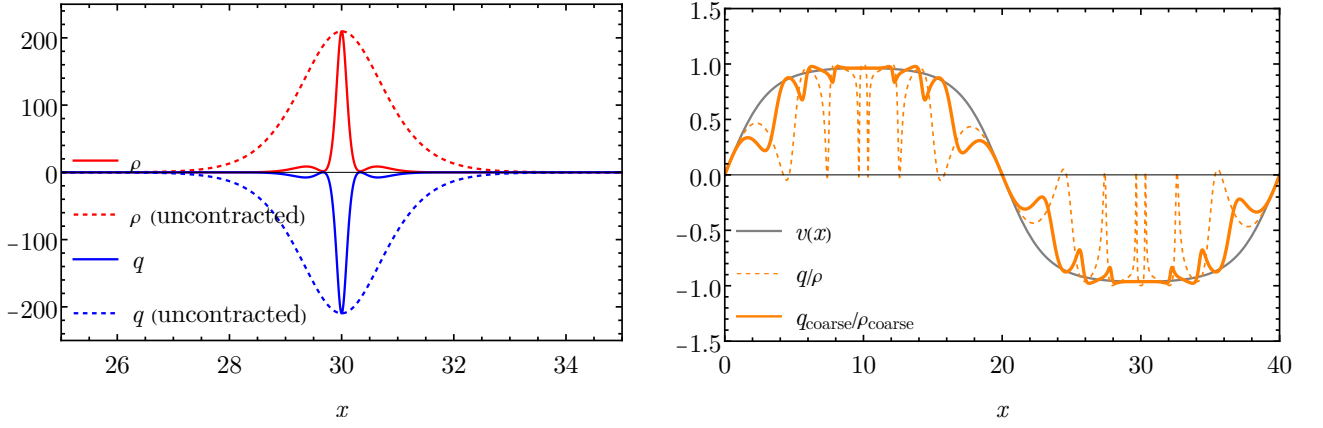

FIG. 1. Information on the stress-energy tensor of the boosted Sine-Gordon field. The left panel gives energy and momentum density plot of one soliton, and compares them to "uncontracted" values computed from a local Lorentz boost on the original stress-energy tensor $T^\mu{}_\nu(0, \boldsymbol{x})$. The right panel gives local velocities $q/\rho$, and compares them to the desired velocity field $v(x) = \tanh^{-1}(-\partial_x \tau(x))$.

One can see that the solid and dashed lines mismatch at all locations but at $x = 30$, at which the peak of $\rho$ and $q$ are the same. This is interpreted as Lorentz contraction: a Lorentz boost contains a coordinate change $x \to \gamma x$, so the energy density profile appears shrunk. The right panel of fig. 1 gives information on the velocity: the gray curve is $v(x) = \tanh(2\sin(2\pi x/L))$, the dashed curve gives $q(x)/\rho(x)$. The solid curve is $q_{\text{coarse}}(x)/\rho_{\text{coarse}}(x)$, which are coarse grained via

$$\rho_{\text{coarse}}(x) = \frac{1}{4}\int_{x-2}^{x+2} \rho(y)\,\mathrm{d}y\,, \quad q_{\text{coarse}}(x) = \frac{1}{4}\int_{x-2}^{x+2} q(y)\,\mathrm{d}y\,. \tag{4}$$

It is clear that while $q/\rho$ have large local oscillations, the coarse grained velocity tracks the desired velocity $v(x)$ closely, indicating that the spatially varying boost has assigned correct velocities to the field.

### B. Relativistic Proca field

The Lagrangian of the Proca field is $\mathcal{L} = -\frac{1}{4}F_{\mu\nu}F^{\mu\nu} - \frac{1}{2}m^2 A_\mu A^\mu$, where $F_{\mu\nu} = \partial_\mu A_\nu - \partial_\nu A_\mu$. Its energy and momentum densities are

$$\begin{aligned} \rho = -T^0{}_0 &= \frac{1}{2}\left[(\partial_t A_i - \partial_i A_t)^2 + m^2(A_i)^2\right] \\ &\quad + \frac{1}{4}(\partial_i A_j - \partial_j A_i)^2 + \frac{1}{2}m^2(A_t)^2 \\ \boldsymbol{q} = T^0{}_i \boldsymbol{e}^i &= -m^2 A_t \boldsymbol{A} - (\partial_t A_j - \partial_j A_t)(\partial_i A_j - \partial_j A_i)\boldsymbol{e}^i\,. \end{aligned} \tag{5}$$

The component $A_t$ is non-dynamical, and can be solved in terms of $\partial_t \boldsymbol{A}$ in Fourier space:

$$A_{t,\boldsymbol{k}} = (-i\boldsymbol{k} \cdot \partial_t \boldsymbol{A}_{\boldsymbol{k}})/(k^2 + m^2)\,. \tag{6}$$

The equation of motion for the dynamical variables is

$$0 = \partial_t^2 \boldsymbol{A} - \nabla^2 \boldsymbol{A} + m^2 \boldsymbol{A}\,. \tag{7}$$

The fields are initialized on a 3D box of side length $L = 19.2m^{-1}$, lattice size $N^3 = 384^3$ and grid size $\Delta x = 0.05m^{-1}$. Periodic boundary conditions apply for this box. We first initialized 3 homogeneous Gaussian random fields $B_i$ with spectrum $\Delta^2_{B_i}(k) = C_1 k^3 \Theta(k_* - k)$, where $k_* = 5m$ (relativistic) and $C_1$ is some constant. We then obtain $A_i$'s by projecting $B_i$'s to the transverse subspace by computing the following in Fourier space:

$$A_{i,\boldsymbol{k}} = (\delta_{ij} - k_i k_j/k^2)B_{j,\boldsymbol{k}}\,. \tag{8}$$

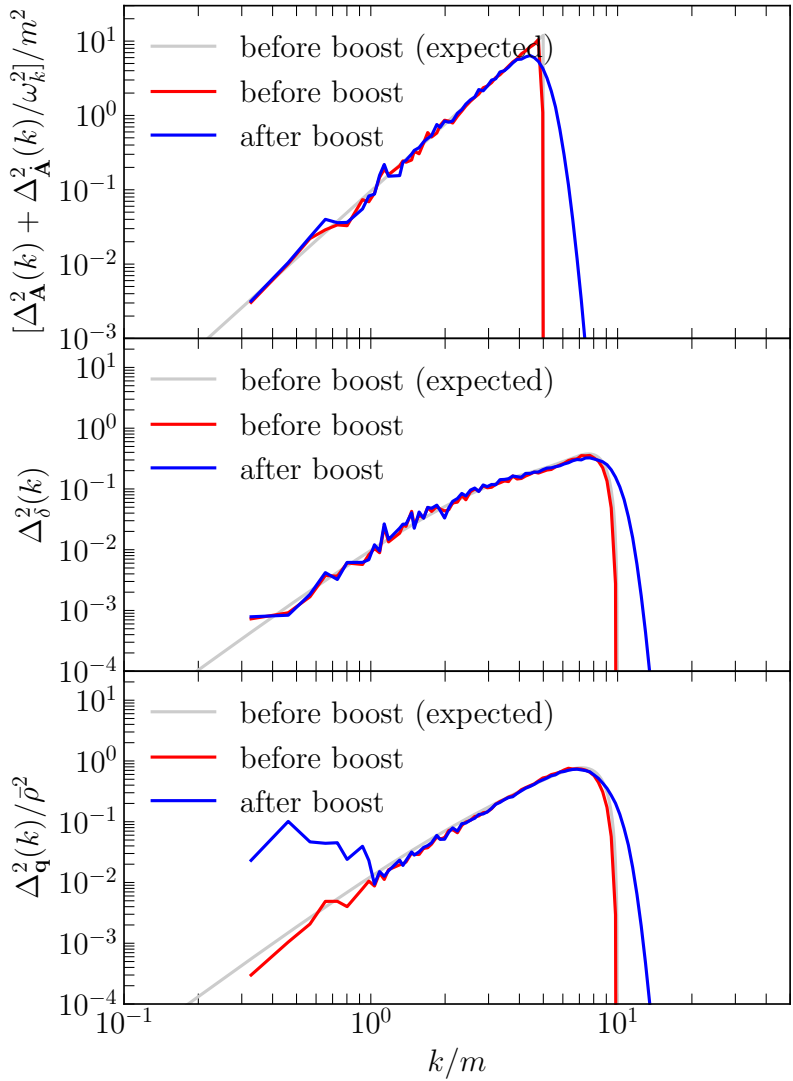

FIG. 2. Spectra of the Proca field, before and after boosting.

The fields $\partial_t A_i$'s were also initialized by projecting $\partial_t B_i$'s, which were generated with spectrum $\Delta^2_{\partial_t B_i}(k) = (k^2 + m^2)\Delta^2_{B_i}(k)$. The time derivative of overdensities $f = \partial_t \delta$ was also generated as a homogeneous Gaussian random field, but with scale-invariant spectrum $\Delta^2_f(k) = C_2 \Theta(k_f - k)$, with $k_f = m$, and with $C_2$ fixed such that $\sqrt{\overline{f^2}} = 0.1$.

Since $v \ll 1$, we have linearized eq. (3) in the main text with respect to $\boldsymbol{v}$. Field $\boldsymbol{A}$ was evolved both forward and backward in $t$ using RK4, and field values at $(\tau(\boldsymbol{x}), \boldsymbol{x})$ were evaluated by interpolating between adjacent time $t$ and $t + \Delta t$ (where $t \leq \tau(\boldsymbol{x}) < t + \Delta t$) at each lattice site $\boldsymbol{x}$. Specifically, we used cubic Hermite interpolation with data $\boldsymbol{A}(t, \boldsymbol{x})$, $\boldsymbol{A}(t + \Delta t, \boldsymbol{x})$, $\partial_t \boldsymbol{A}(t, \boldsymbol{x})$, $\partial_t \boldsymbol{A}(t + \Delta t, \boldsymbol{x})$, and did similar interpolations for quantities like $\nabla \cdot \boldsymbol{A}$.

See fig. 2 for the spectra before and after boost. The top panel gives the field spectrum, which is cut-off at $k_*/m = 5$. The middle and bottom panel give the spectra for energy density contrast $\delta = \rho/\bar{\rho} - 1$ and normalized momentum density $\boldsymbol{q}/\bar{\rho}$. One can see that, at low-$k$, the field (top) and overdensity (middle) spectra are changed negligibly by the boost, whereas the momentum spectrum (bottom) gains a large contribution with cutoff at $k_\delta/m = 1$, as expected. Note that the momentum spectrum is not zero even before boosting due to small scale field fluctuations. At high-$k$, all 3 spectra gains higher momentum modes $k > k_*$ from the boost. Consider a wavepacket in the field: its speed is $v_* = k_*/\sqrt{k_*^2 + m^2} = 0.98$, so under the boost, the max possible speed of the wavepacket becomes $v' = (v_* + v)/(1 + v_* v)$, where $v \approx 0.2$. The boosted momenta becomes $mv'/\sqrt{1 - (v')^2} \approx 6m$, which is consistent with the figure.

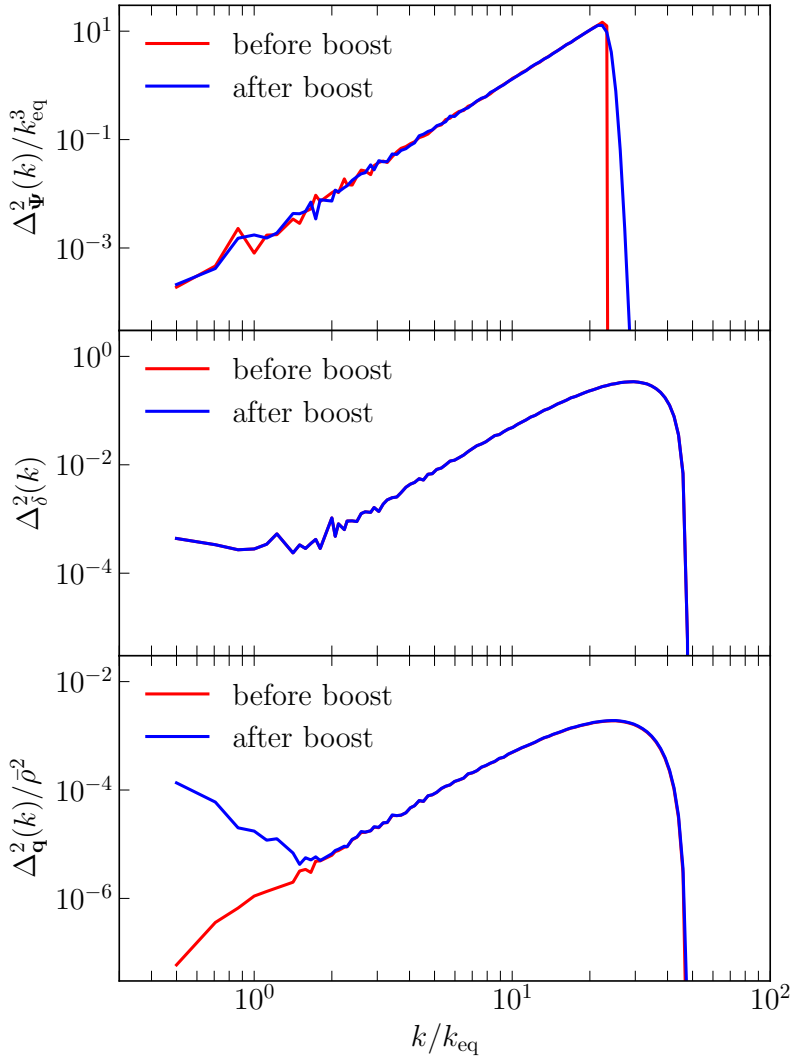

FIG. 3. Spectra of the Schrödinger-Poisson field, before and after boosting.

### C. Spin-1 wave dark matter

The fields are initialized on a 3D box of (comoving) side length $L = 12.566 k_{\rm eq}^{-1}$ and lattice size $N^3 = 384^3$. Periodic boundary conditions apply for this box. The time at which the field is initialized is $t_1$. The smallest and largest wavenumbers associated with the box are $k_{\rm IR} = 2\pi/L$ and $k_{\rm UV} = k_{\rm IR} N/2$. At $t_1$, the scale factor is $\frac{a(t_1)}{a_{\rm eq}} = 16$, the Hubble parameter is $H(t_1) = 0.015625 H_{\rm eq}$. Some relevant numbers are given by:

$$\frac{a(t_1) H(t_1)}{k_{\rm eq}} = 0.25, \quad \frac{k_{\rm IR}}{k_{\rm eq}} = 0.5, \quad \frac{k_*}{k_{\rm eq}} = 24, \quad \frac{k_{\rm UV}}{k_{\rm eq}} = 96, \quad m = 10 H_{\rm eq}, \quad v_* \equiv \frac{k_*}{a(t_1) m} = 0.15 \,. \qquad (9)$$

We have chosen unrealistic parameters for $m$ for ease of visualization.

For $\mathcal{R}_{\bm{k}}$, we generated a scale invariant homogeneous Gaussian random field with amplitude $\mathcal{A}_s = 2 \times 10^{-6}$. Note that we have enhanced the value of $\mathcal{A}_s$ compared to the observed values for ease of visualization. We then computed $\Phi$, $\partial_t \Phi$, $\delta$ and $\partial_t \delta$ at $t_1$ by scaling $\mathcal{R}_{\bm{k}}$ in Fourier space with the appropriate kernels. For $\bm{\Psi}$, we generated 6 inhomogeneous Gaussian random fields with spectrum $\Delta^2(k) = C_1 k^3 \Theta(k_* - k)$ ($k_* = 24 k_{\rm eq}$) and variance perturbation $f_{\bm{\Psi}} = \delta + \Phi$, and set them to be the real and imaginary part of the 3 components of $\bm{\Psi}$.

The spectra for the Schrödinger-Poisson field is given in fig. 3. One can see that the overdensity spectrum has scale invariant large scale perturbations due to inhomogeneous Gaussian random field. Moreover, the boost have negligible impact on the field and overdensity spectrum, and it enhances large scale momentum perturbations, as expected.



\end{document}